\documentstyle[12pt]{article}

\renewcommand{\baselinestretch}{1.2}
\newlength{\dinwidth}
\newlength{\dinmargin}
\def\draftdate{\relax}
\def\mda{\relax}
\def\mua{\relax}
\def\mla{\relax}
\def\draft{
\def\thtystars{******************************}
\def\sixtystars{\thtystars\thtystars}
\typeout{}
\typeout{\sixtystars**}
\typeout{* Draft mode!
         For final version remove \protect\draft\space in source file *}
\typeout{\sixtystars**}
\typeout{}
\def\draftdate{\today}
\def\mua{\marginpar[\boldmath\hfil$\uparrow$]%
                   {\boldmath$\uparrow$\hfil}%
                    \typeout{marginpar: $\uparrow$}\ignorespaces}
\def\mda{\marginpar[\boldmath\hfil$\downarrow$]%
                   {\boldmath$\downarrow$\hfil}%
                    \typeout{marginpar: $\downarrow$}\ignorespaces}
\def\mla{\marginpar[\boldmath\hfil$\rightarrow$]%
                   {\boldmath$\leftarrow $\hfil}%
                    \typeout{marginpar: $\leftrightarrow$}\ignorespaces}
\overfullrule 5pt
\oddsidemargin -15mm
\marginparwidth 29mm
}
\begin{document}
\newcommand{\beq}{\begin{equation}}
\newcommand{\eeq}{\end{equation}}
\newcommand{\bea}{\begin{eqnarray}}
\newcommand{\eea}{\end{eqnarray}}
\newcommand{\ba}{\begin{array}}
\newcommand{\ea}{\end{array}}
\newcommand{\ds}{\displaystyle}
\newcommand{\noi}{\noindent}
\newcommand{\tfrac}[2]{\frac{\textstyle #1}{\textstyle #2}}
\unitlength1cm
\setcounter{page}{0}
\thispagestyle{empty}
\begin{flushright}
BI-TP 95/25 \\
hep-ph/9507204 \\
June 1995
\end{flushright}

\vspace*{\fill}
\begin{center}
\begin{LARGE}
\begin{bf}
Luminosities for Vector-Boson \\
Vector-Boson Scattering \\
at High Energy Colliders \\
\end{bf}
\end{LARGE}

\vspace{2cm}

\begin{large}
I.\ Kuss\footnote{\normalsize e-mail: kuss@hrz.uni-bielefeld.de},
H.\ Spiesberger \\
\end{large}

\vspace{1cm}

Universit\"at Bielefeld\\
Fakult\"at f\"ur Physik\\
Postfach 10 01 31\\
D--33501 Bielefeld, Germany\\

\vspace{2cm}

\begin{abstract}
\noindent
We derive exact expressions for luminosities of massive vector-boson
pairs which can be used to describe the cross sections for processes in
hadron collisions or $e^+e^-$ annihilation which proceed via
two-vector-boson scattering. Our approach correctly takes into account
the mutual influence of the emission of one vector boson on the
emission of a second one. We show that only approximately the exact
luminosities can be factorized into convolutions of single-vector-boson
distributions. Numerical results are given and compared to simplified
approaches.
\end{abstract}

\end{center}

\vspace*{\fill}

\newpage


\section{Introduction}

Hadron colliders will produce the electroweak vector bosons $W^{\pm}$
and $Z$ with a high rate at large energies, and many processes, like
e.g.\ Higgs-boson production or heavy-quark production, can proceed
via vector-boson vector-boson scattering. The experimental study of
these and similar processes is expected to lead to an understanding of
the Higgs sector of the electroweak standard model and eventually of
the electroweak symmetry breaking mechanism. In addition, vector-boson
pair production at hadron colliders will provide information on the
self-couplings of the $W^{\pm}$ and $Z$ bosons and possibly play an
important role in the search for new physics.

In lowest order, vector bosons can be produced by quark-antiquark
annihilation in hadron collisions. However, at high energies,
higher-order processes where vector bosons emitted from incoming quarks
or antiquarks initiate a hard scattering process can be enhanced by
logarithmic factors and thus can compete with the lowest-order
production mechanism. These processes have successfully been described
with the help of the effective vector boson method (EVBM) which
applies the concept of partons in a hadron to the case of vector
bosons: vector bosons are viewed as partons in quarks and electrons,
as quarks and gluons are partons in hadrons. In analogy to the
Weizs\"acker-Williams approximation of QED \cite{WWA} the cross
section for a scattering process $a + A \rightarrow X$ at a
center-of-mass energy $s$ is factorized into probability densities
$P_{pol}^{V/a}(z)$ for finding a vector boson $V$ with polarization
$pol$ in the incoming fermion $a$, and hard vector-boson scattering
cross sections at a reduced center-of-mass energy $xs$:
\begin{equation}
d\sigma(a+A \rightarrow X, s) =
\int\limits_{x_{\mathrm{min}}}^{1} \,dx
\sum_V \sum_{pol} P^{V/a}_{pol}(x)
d\sigma(V_{pol} + A \rightarrow X,xs).
\end{equation}
The basic assumptions in the effective vector boson method are that the
dominant contributions for producing the final state $X$ is due to
vector-boson initiated processes and that the cross section for the
scattering of an off-shell vector boson can be related to the
corresponding on-shell cross section.

In the application of the method to processes with two intermediate
vector bosons (see Fig.\ 1) it was assumed that convolutions of
single-vector-boson probability densities are sufficient to obtain
luminosities for vector-boson pairs,
\begin{equation}
{\cal L}^{V_1 V_2/ab}_{pol_1 pol_2}(x) =
\int\limits_{z_{\mathrm{min}}}^1 \frac{dz}{z} P^{V_1/a}_{pol_1}(z)
 P^{V_2/b}_{pol_2}(x/z),
\label{deflumi}
\end{equation}
which can be used to express the cross section for two-fermion
scattering in terms of the vector-boson vector-boson scattering cross
section:
\begin{equation}
d\sigma(a+b \rightarrow X, s) =
\int\limits_{x_{\mathrm{min}}}^1 \,dx\sum_{V_1,V_2}\sum_{pol_1,pol_2}
{\cal L}_{pol_1 pol_2}^{V_1 V_2/ab}(x)
d\sigma(V_1^{pol_1} + V_2^{pol_2} \rightarrow X, xs).
\end{equation}

\begin{figure}
\unitlength1cm
\begin{center}
\begin{picture}(4,6)
\put(-.2,5.3){$1(l_1)$}
\put(-.2,0.8){$2(l_2)$}
\put(3.3,5.3){$1'(p_1)$}
\put(3.3,0.8){$2'(p_2)$}
\put(4.,3.0){$W(p_W),{\cal W}$}
\put(-.1,4.2){$V_1(q_1),M_1$}
\put(-.1,2.0){$V_2(q_2),M_2$}
\put(-6.5,-8.0){\includegraphics{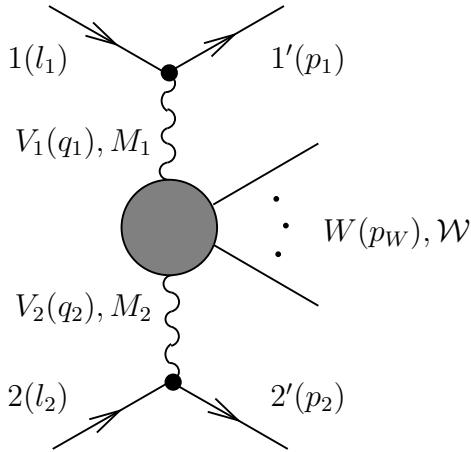}}
\end{picture}
\end{center}
\caption{The vector-boson scattering diagram}
\label{fig1}
\end{figure}

The possibility to generalize the equivalent photon approximation to
the case of massive vector bosons was first noted in \cite{kane1} and
explicitly formulated in \cite{dawson,kane2,lindfors}. Originally, the
concept was invented for the description of processes at very high
energies and thus included a number of approximations valid at high
energies only. These approximations were partly of kinematic origin
and concerned the neglect of mass terms, or of the transverse momentum
of the intermediate vector bosons. According to the details of the
approximation, a variety of versions for vector-boson distributions
with differing numerical results can be found in the literature. The
most simple of these approximations---the leading logarithmic
approximation (LLA)---amounts to taking a zero-mass limit. In addition,
since it was observed that for the production of a heavy Higgs particle
in vector-boson scattering the cross section is dominated by
longitudinal polarization \cite{cahn}, first applications of the method
neglected contributions from transversely polarized vector bosons and
the interference between amplitudes for different polarizations.

Comparisons with exact calculations have shown that the method is
indeed helpful and leads to reliable results, in particular for Higgs
boson production \cite{HIGGS} and for heavy fermion production
\cite{HEAVYQ}. The application to vector-boson vector-boson scattering
off the Higgs resonance \cite{VVscatt} was less successful; the
effective vector boson method overestimated the exact result
\cite{dicus1}. Adding the (positive) contribution from transversely
polarized vector bosons \cite{TRANSV} could, of course, not lead to an
improved agreement between the EVBM and exact calculations for
vector-boson pair production.

In \cite{tung1} it was shown that approximations of kinematic origin
can be avoided and a set of exact vector-boson distributions was
derived. There it was also shown that interference terms (i.e.\
non-diagonal contributions) do not appear in the case of
single-vector-boson processes (see also \cite{lindfors}). The only
remaining necessary assumption in using the EVBM for
single-vector-boson processes concerned the off-shell behaviour of the
hard scattering cross section.

We will show that the improvement obtained with the results of
\cite{tung1} is not sufficient to accurately describe two-vector-boson
processes. The simple convolution of two single-vector-boson
probabilities\footnote{Explicit expressions for the luminosity
    functions derived from convolutions of single-boson distributions
    in the leading logarithmic approximation have been given in
    \cite{lindfors3,renard} and the last but one reference of
    \cite{VVscatt}.}
as in Eq.\ (\ref{deflumi}) ignores the mutual influence of the emission
of one boson on the probability for the emission of another. In
addition, interference contributions need not vanish, as has been
noticed in the specific example of Higgs boson production in
\cite{dobrov}. This is in analogy to two-photon processes \cite{budnev}
where it was shown already in \cite{bonneau} that the extension of the
Weizs\"acker-Williams method from one photon to the case of two photons
is not straightforward.

The main purpose of the present work is the extension of the effective
vector boson method to the case of processes with two vector bosons,
as needed in the study of vector-boson vector-boson scattering.
It thus combines the exact treatment of the two-boson kinematics,
presented for photons in \cite{bonneau}, with the exact definition of
vector-boson distributions, presented for single vector bosons in
\cite{tung1}. Our derivation (section 2) will not use any kinematic
approximation. It turns out that non-diagonal terms are indeed needed.
In section 3, we will present exact luminosity functions for
vector-boson pairs in quark or electron initiated processes. One can
then identify the additional approximation needed to reduce these
luminosities to convolutions of the exact single-vector-boson densities
of \cite{tung1} (section 4) and in the high energy limit we also
recover the leading logarithmic versions of vector-boson distributions
as used in the literature (section 5). Finally, in section 6 we will
also present numerical results for these exact luminosities and compare
them with the ones of simplified approaches.

Despite of the fact that both single-vector-boson distributions and
two-vector-boson luminosities can be obtained exactly without any
approximation, there remains the question whether the set of Feynman
diagrams that can be described with the help of the EVBM is indeed the
dominating one. The answer to this problem depends on the process and
has to be found in a case-by-case study. Of particular concern in this
respect is the question whether the considered subset of diagrams is
gauge invariant. In \cite{kleiss} it was observed that for off-shell
vector-boson scattering there may occur strong gauge cancellations
between those contributions taken into account in the EVBM and
bremsstrahlung diagrams which are ignored. Motivated by this, Kunszt
and Soper \cite{kunszt} argued that the extrapolation to off-shell
masses is not always guaranteed, but for heavy Higgs-boson production
they show in an axial gauge the validity of the basic assumption that
the extrapolation to off-shell masses is indeed a smooth one.

Our final explicit expressions for the two-vector-boson luminosities are
obtained with specific simple assumptions for the off-shell behaviour
of the vector-boson scattering cross sections. However, we keep a clean
separation of exactly calculable parts and model assumptions and our
expressions are written in a form which allows for an easy accommodation
of an improved off-shell dependence, as soon as the corresponding
information would be available. Apart from these caveats, our
luminosities are exact results of a calculation of a subset of Feynman
diagrams. In particular, their range of validity is not restricted to
large energies. Therefore we will also present some of the results for
an energy of $\sqrt{s} = 500$ GeV, relevant for a next-generation
$e^+e^-$ collider. The alternative approach of using convolutions
of the LLA single-vector-boson distributions is not applicable at these
small energies.

\section{General Formalism of the Effective Vector Boson Method}

We consider the production of an arbitrary state $W$ in the 2-fermion
scattering process (see Fig.\ 1)
\begin{equation}
1(l_1) + 2(l_2) \rightarrow 1'(p_1) + 2'(p_2) + W(p_W).
\label{process}
\end{equation}
The 4-momenta of the incoming and outgoing fermions are denoted by
$l_1$, $l_2$ and $p_1$, $p_2$, resp., the total center-of-mass energy
squared is given by $s = (l_1 + l_2)^2$. The final state $W$, which may
contain any number of particles, has 4-momentum $p_W$ and its invariant
mass squared will be denoted by ${\cal W}^2=p_W^2$. The cross section
for the process (\ref{process}) is given by
\begin{equation}
\sigma_{ff}=\frac{1}{2s}\frac{1}{(2\pi)^2}\int \frac{d^3p_1}{2p_1^0}
\int\frac{d^3p_2}{2p_2^0}
\int d\rho_W\overline{{|{\cal M}_{ff}|}^2}
\delta^{(4)}(l_1+l_2-p_1-p_2-p_W).\label{sigma0}
\end{equation}
In (\ref{sigma0}), $\overline{{|{\cal M}_{ff}|}^2}$ is the squared
amplitude for the two-fermion initiated process, averaged and summed
over helicities and $ d\rho_W$ is the phase space element for the state
$W$.

For high energies one can neglect the fermion masses. With the help of
the momentum transfers $q_j = l_j - p_j$, $j=1, 2$ and using the
dimensionless variables
\begin{equation}
x = \frac{{\cal W}^2}{s}, ~~~~~
z = \frac{M_X^2}{s}, ~~~~{\rm with} ~~ M_X^2 = (p_W + p_2)^2,
\end{equation}
as well as
\begin{equation}
Q^2_2=\frac{1}{1-\tfrac{q_1^2}{M_X^2}}q_2^2,
\label{bigq2}
\end{equation}
one can parametrize the phase space by
\begin{eqnarray}
\sigma_{ff} &=& \frac{1}{32s} \int\limits_{x_0}^1dx
                              \int\limits_{x}^1\frac{dz}{z}
\int\limits_{-s(1-z)}^0\!\!\!\!\!\!dq_1^2
\int\limits_{-sz(1-\frac{x}{z})}^0\!\!\!\!\!\!\!dQ^2_2
\int\limits_0^{2\pi}\frac{d\varphi_1}{2\pi}
\int\limits_0^{2\pi}\frac{d\varphi_2}{2\pi}\nonumber\\
            & & \int d\rho_W\overline{{|{\cal M}_{ff}|}^2}
\delta^{(4)}(l_1+l_2-p_1-p_2-p_W).\label{sigma1}
\end{eqnarray}
Here, $x_0 = {\cal W}^2_{min}/s$ is the minimal value of the invariant
mass squared of the final state $W$ normalized to the total
center-of-mass energy. In case of an $n$-particle final state,
${\cal W}_{min}$ is equal to the sum of the masses of these particles.
$\varphi_1$ and $\varphi_2$ are azimuthal angles for the momenta $p_1$
and $p_2$, resp., defined in Breit systems $B_1$ and $B_2$ in which
either $q_1$ or $q_2$ has only a non-vanishing $z$-component (see
Appendix \ref{app1}).

If the process (\ref{process}) proceeds via the vector-boson fusion
mechanism as shown in Fig.\ (\ref{fig1}), the expression for the
amplitude ${\cal M}_{ff}$ is given by\footnote{A sum must be taken over
                all vector-boson pairs $V_1,V_2$ which can couple to
                the fermions and produce the final state $W$. We do not
                treat the interference terms here, but the extension of
                our formalism to take them into account is
                straightforward.}
\begin{equation}
{\cal M}_{ff} = e^2\sum_{m,n=-1}^{1}(-1)^{m+n}
                \frac{j_1(l_1,p_1)\cdot\epsilon^\ast_{1}(m)}
                {q_1^2-M_1^2}\frac{
                j_2(l_2,p_2)\cdot\epsilon^\ast_{2}(n)}{q_2^2-M_2^2}
                {\cal M}(m,n),
\label{amp}
\end{equation}
where the $\epsilon_j(m)$ are polarization vectors for the vector
boson $V_j$ with mass $M_j$ and helicity $m=0,\pm 1$ in the
center-of-mass system $C$ of $q_1 + q_2$. Explicit expressions for them
are given in App.\ \ref{app2}. The $j_j(l_j,p_j)$ are fermionic current
4-vectors, $e$ is the positron charge, and ${\cal M}(m,n)$ is the
amplitude for the production of the final state $W$ from vector bosons
$V_1$ and $V_2$ with helicities $m$ and $n$, resp. The amplitudes
${\cal M}(m,n)$ must be evaluated at off-shell values of $q_1^2$ and
$q_2^2$. The polarization vectors are normalized according to
\begin{equation}
\epsilon_j(m)\cdot\epsilon_j^\ast(m')=\delta_{m,m'}(-1)^m,\qquad
j=1,2\;,
\end{equation}
and satisfy the completeness relation
\begin{equation}
\sum_{m=-1}^1\epsilon_j^\mu(m)\epsilon_j^{\ast\nu}(m)=
-g^{\mu\nu}+\frac{q_j^{\mu} q_j^{\nu}}{M_j^2},
\qquad\mbox{(no sum on $j$)}\;,
\end{equation}
which corresponds to writing the vector-boson propagators in the unitary
gauge.

The expression for the squared amplitude, averaged over the spin states
of the initial fermions and summed over the spins of the final state
fermions is
\begin{equation}
\overline{{|{\cal M}_{ff}|}^2}=4\,e^4\sum_{m,m'=-1}^1\sum_{n,n'=-1}^1
(-1)^{m+m'+n+n'}\frac{\tilde{T_1}(m,m')}{(q_1^2-M_1^2)^2}
\frac{\tilde{T_2}(n,n')}{(q_2^2-M_2^2)^2}
{\cal M}(m,n){\cal M}^\ast(m',n'),
\label{m2}
\end{equation}
with the fermionic tensors
\begin{equation}
\tilde{T_j}(m,m')=\frac{1}{4}\sum_{pol}j_j(l_j,p_j)\cdot
\epsilon_j^\ast(m) j_j^\ast(l_j,p_j)\cdot\epsilon_j(m').
\end{equation}
The tensor $\tilde{T_j}(m,m')$ can be decomposed into two parts with
different combinations of the vector and axial-vector coupling
constants $v_j$ and $a_j$ of the vector bosons $V_j$ by the relation
\begin{equation}
\tilde{T}_j(m,m')=(v_j^2+a_j^2)\tilde{\cal C}_j(m,m')
+2v_ja_j\tilde{\cal S}_j(m,m').
\end{equation}
The tensors $\tilde{\cal C}_j(m,m')$ and $\tilde{\cal S}_j(m,m')$
(i.e.\ tensors in helicity space) are given by
\begin{equation}
\tilde{\cal C}_j(m,m') =
  p_j\cdot\epsilon_j^\ast(m)l_j\cdot\epsilon_j(m')
+ p_j\cdot\epsilon_j(m')l_j\cdot\epsilon_j^\ast(m)
- p_j\cdot l_j\epsilon_j^\ast(m)\cdot\epsilon_j(m')
\label{Cexp}
\end{equation}
and
\begin{equation}
\tilde{\cal S}_j(m,m')=i\epsilon_{\alpha\beta\gamma\delta}p_j^\alpha
{\epsilon_j^{\beta}}^{\ast}(m) l_j^\gamma \epsilon_j^\delta(m'),
\label{Sexp}
\end{equation}
with $\epsilon_{0123} =1$.

Factorizing the $\varphi_2$-dependence of the tensor components of
$\tilde{\cal C}_j(m,m')$ and $\tilde{\cal S}_j(m,m')$ which is given
in terms of simple exponential functions, we define
$\varphi_2$-independent tensors ${\cal C}_j(m,m')$ and
${\cal S}_j(m,m')$:
\begin{eqnarray}
\tilde{\cal C}_1(m,m')&=&{\cal C}_1(m,m')e^{i(m-m')\varphi_2},\cr
\tilde{\cal S}_1(m,m')&=&{\cal S}_1(m,m')e^{i(m-m')\varphi_2},\cr
\tilde{\cal C}_2(n,n')&=&{\cal C}_2(n,n')e^{-i(n-n')\varphi_2},\cr
\tilde{\cal S}_2(n,n')&=&{\cal S}_2(n,n')e^{-i(n-n')\varphi_2},
\label{varphi}
\label{Tdef}
\end{eqnarray}
for which the following relations hold:
\begin{eqnarray}
{\cal C}_j(m',m)&=&{\cal C}_j^\ast(m,m'),\cr
{\cal S}_j(m',m)&=&{\cal S}_j^\ast(m,m'),\cr
{\cal C}_j(-m',-m)&=&(-1)^{m+m'}{\cal C}_j(m,m'),\cr
{\cal S}_j(-m',-m)&=&-(-1)^{m+m'}{\cal S}_j(m,m').\label{CSrel}
\end{eqnarray}
The last relation in (\ref{CSrel}) implies
\begin{eqnarray}
{\cal S}_j(+-) & = & 0, \quad\mbox{and}\cr
{\cal S}_j(00) & = & 0.
\end{eqnarray}
Consequently, ${\cal C}_j(++)$, ${\cal C}_j(00)$, ${\cal C}_j(+-)$
and ${\cal C}_j(+0)$ can be chosen as the $2\times4$ independent
components of ${\cal C}_j(m,m')$ and the ${\cal S}_j(m,m')$ have the
$2\times 2$ independent components ${\cal S}_j(++)$ and ${\cal
S}_j(+0)$. We illustrate this situation by writing down $\tilde{{\cal
C}}_1(m,m')$ and $\tilde{{\cal S}}_1(m,m')$ in matrix form:
\begin{equation}
\arraycolsep0cm
\tilde{{\cal C}}_1(m,m')=\left(\begin{array}{lrlrl}
{\cal C}_1(++)&&{\cal C}_1^\ast(+0)e^{-i\varphi_2}&&
{\cal C}_1^\ast(+-)e^{-2i\varphi_2}\cr
{\cal C}_1(+0)e^{i\varphi_2}&&{\cal C}_1(00)&-&{\cal
 C}_1^\ast(+0)e^{-i\varphi_2}\cr
{\cal C}_1(+-)e^{2i\varphi_2} & \,- & {\cal C}_1(+0)e^{i\varphi_2}
 &&{\cal C}_1(++)
\end{array}\right)
\end{equation}
and
\begin{equation}
\arraycolsep0cm
\tilde{{\cal S}}_1(m,m')=\left(\begin{array}{lrlrl}
{\cal S}_1(++) && {\cal S}_1^\ast(+0)e^{-i\varphi_2} && ~~~~~0 \cr
{\cal S}_1(+0)e^{i\varphi_2} && ~~~~~0 &&
                        {\cal S}_1^\ast(+0)e^{-i\varphi_2}\cr
{}~~~~~0 && {\cal S}_1(+0)e^{i\varphi_2} &\,-& {\cal S}_1(++)
\end{array}\right),
\end{equation}
where the columns from left to right correspond to $m=+,0,-$ and the
rows from top to bottom to $m'=+,0,-$. Expressions for the independent
components in terms of the integration variables in (\ref{sigma1}) are
given in Appendix \ref{appB}. Similar decompositions can be given for
$\tilde{{\cal C}}_2(m,m')$ and $\tilde{{\cal S}}_2(m,m')$. The
quantities ${\cal C}_2(m,m')$ and ${\cal S}_2(m,m')$ turn out to be
real.

Carrying out the integration over $\varphi_2$, there remain altogether
19 terms in the $m$, $m'$, $n$, $n'$ helicity space, out of which nine
have $h=m-m'=n-n'=0$ (they are diagonal in the helicities of $V_1$ and
$V_2$), four have $h=1$, four have $h=-1$ and the other two have $h=2$
and $h=-2$, resp. For the case of two-photon interactions this
classification has been given in \cite{bonneau}. Using this
decomposition, one can write the expression in
Eq.\ (\ref{m2}) in the following way:
\begin{eqnarray} &&
\frac{1}{2\pi}\int_0^{2\pi}d\varphi_2\sum_{m,m',n,n'=-1}^{1}
(-1)^{m+m'+n+n'}\tilde{T}_1(m,m')\tilde{T}_2(n,n')
{\cal M}(m,n){\cal M}^\ast(m',n')
\nonumber
\end{eqnarray}
\begin{eqnarray}
&=&(v_1^2+a_1^2)(v_2^2+a_2^2)
   \left(K_{TT}M_{TT} + K_{TL}M_{TL} + K_{LT}M_{LT}
         + K_{LL}M_{LL}\right.\nonumber \\
&&\left. + K_{TLTL}M_{TLTL}+K_{TTTT}M_{TTTT}
-K^{Im}_{TTTT}M^{Im}_{TTTT}-K^{Im}_{TLTL}M^{Im}_{TLTL}\right)\\
&&+(2v_1a_1)(2v_2a_2) \left(K_{\overline{TT}}M_{\overline{TT}}
+K_{\overline{T}L\overline{T}L}M_{\overline{T}L\overline{T}L}
-K^{Im}_{\overline{T}L\overline{T}L}
 M^{Im}_{\overline{T}L\overline{T}L}\right)\nonumber\\
&&+(v_1^2+a_1^2)(2v_2a_2) \left(K_{T\overline{T}}M_{T\overline{T}}
+K_{L\overline{T}}M_{L\overline{T}}+K_{TL\overline{T}L}
 M_{TL\overline{T}L}
-K^{Im}_{TL\overline{T}L}M^{Im}_{TL\overline{T}L}\right)\nonumber\\
&&+(2v_1a_1)(v_2^2+a_2^2) \left(K_{\overline{T}T}M_{\overline{T}T}
+K_{\overline{T}L}M_{\overline{T}L}+K_{\overline{T}LTL}
 M_{\overline{T}LTL}
-K^{Im}_{\overline{T}LTL}M^{Im}_{\overline{T}LTL}\right)
\nonumber \\[1em]
&=&\sum_{pol}c_{f,pol}\,K_{pol}\,M_{pol},
\label{terms19}
\end{eqnarray}
where the last line defines the notation to be used below, with
$pol$ being labels for the polarizations, $pol = TT$, $\overline{TT}$,
etc. $c_{f,pol}$ contain the fermionic coupling constants
and---depending on the index $pol$---can take the values
$c_{f,pol}=(v_1^2+a_1^2)(v_2^2+a_2^2)$, $(2v_1a_1)(2v_2a_2)$,
$(v_1^2+a_1^2)(2v_2a_2)$ and $(2v_1a_1)(v_2^2+a_2^2)$. The quantities
$K_{pol}$, which are five-fold differential luminosities---they depend
on ${\cal W}^2,q_1^2,q_2^2,M_X^2$ and $\varphi_1$---are defined by
\begin{eqnarray}
K_{TT}&=&4{\cal C}_1(++){\cal C}_2(++),\cr
K_{\overline{TT}}&=&4{\cal S}_1(++){\cal S}_2(++),\cr
K_{TL}&=&2{\cal C}_1(++){\cal C}_2(00),\cr
K_{LT}&=&2{\cal C}_1(00){\cal C}_2(++),\cr
K_{LL}&=&{\cal C}_1(00){\cal C}_2(00),\cr
K_{TLTL}&=&8{\it Re}[{\cal C}_1(+0)]{\cal C}_2(+0),\cr
K_{\overline{T}L\overline{T}L}&=&8{\it Re}[{\cal S}_1(+0)]
{\cal S}_2(+0),\cr
K_{TTTT}&=&2{\it Re}[{\cal C}_1(+-)]{\cal C}_2(+-),\cr
K_{T\overline{T}}&=&4{\cal C}_1(++){\cal S}_2(++),\cr
K_{\overline{T}T}&=&4{\cal S}_1(++){\cal C}_2(++),\cr
K_{\overline{T}L}&=&2{\cal S}_1(++){\cal C}_2(00),\cr
K_{L\overline{T}}&=&2{\cal C}_1(00){\cal S}_2(++),\cr
K_{TL\overline{T}L}&=&8{\it{Re}}[{\cal C}_1(+0)]{\cal S}_2(+0),\cr
K_{\overline{T}LTL}&=&8{\it Re}[{\cal S}_1(+0)]{\cal C}_2(+0),\cr
K^{Im}_{TL\overline{T}L}&=&8{\it Im}[{\cal C}_1(+0)]{\cal S}_2(+0),\cr
K^{Im}_{\overline{T}LTL}&=&8{\it Im}[{\cal S}_1(+0)]{\cal C}_2(+0),\cr
K^{Im}_{TLTL}&=&8{\it Im}[{\cal C}_1(+0)]{\cal C}_2(+0),\cr
K^{Im}_{\overline{T}L\overline{T}L}&=&
                          8{\it Im}[{\cal S}_1(+0)]{\cal S}_2(+0),\cr
K^{Im}_{TTTT}&=&2{\it Im}[{\cal C}_1(+-)]{\cal C}_2(+-),
\label{Kdef}
\end{eqnarray}
with ${\cal C}_j(m,m')$ and ${\cal S}_j(m,m')$ from (\ref{cs}) and
(\ref{csprime}) (see App.\ \ref{appB}). The averaged sums of products
of amplitudes for the vector-boson scattering processes, $M_{pol}$, to
be simply called squared amplitudes in what follows, are defined through
\begin{eqnarray}
M_{TT}&=&\frac{1}{4}( {|{\cal M}(++)|}^2+{|{\cal M}(--)|}^2
+{|{\cal M}(+-)|}^2+{|{\cal M}(-+)|}^2 ),\cr
M_{\overline{TT}}&=&\frac{1}{4}( {|{\cal M}(++)|}^2+{|{\cal M}(--)|}^2
-{|{\cal M}(+-)|}^2-{|{\cal M}(-+)|}^2 ),\cr
M_{TL}&=&\frac{1}{2}( {|{\cal M}(+0)|}^2+{|{\cal M}(-0)|}^2),\cr
M_{LT}&=&\frac{1}{2}( {|{\cal M}(0+)|}^2+{|{\cal M}(0-)|}^2),\cr
M_{LL}&=&{|{\cal M}(00)|}^2,\cr
M_{TLTL}&=&\frac{1}{4}{\it Re}[{\cal M}(++){\cal M}^\ast(00)
+{\cal M}(--){\cal M}^\ast(00)-{\cal M}(+0){\cal M}^\ast(0-)
-{\cal M}(-0){\cal M}^\ast(0+)],\cr
M_{\overline{T}L\overline{T}L}&=&\frac{1}{4}{\it Re}[{\cal M}(++){\cal
 M}^\ast(00)
+{\cal M}(--){\cal M}^\ast(00)+{\cal M}(+0){\cal M}^\ast(0-)
+{\cal M}(-0){\cal M}^\ast(0+)],\cr
M_{TTTT}&=&{\it Re}[{\cal M}(++){\cal M}^\ast(--)],\cr
M_{T\overline{T}}&=&\frac{1}{4}( {|{\cal M}(++)|}^2-{|{\cal M}(--)|}^2
-{|{\cal M}(+-)|}^2+{|{\cal M}(-+)|}^2 ),\cr
M_{\overline{T}T}&=&\frac{1}{4}( {|{\cal M}(++)|}^2-{|{\cal M}(--)|}^2
+{|{\cal M}(+-)|}^2-{|{\cal M}(-+)|}^2 ),\cr
M_{\overline{T}L}&=&\frac{1}{2}( {|{\cal M}(+0)|}^2-{|{\cal
 M}(-0)|}^2),\cr
M_{L\overline{T}}&=&\frac{1}{2}( {|{\cal M}(0+)|}^2-{|{\cal
 M}(0-)|}^2),\cr
M_{TL\overline{T}L}&=&\frac{1}{4}{\it Re}[{\cal M}(++){\cal
 M}^\ast(00)
-{\cal M}(--){\cal M}^\ast(00)+{\cal M}(+0){\cal M}^\ast(0-)
-{\cal M}(-0){\cal M}^\ast(0+)],\cr
M_{\overline{T}LTL}&=&\frac{1}{4}{\it Re}[{\cal M}(++){\cal
 M}^\ast(00)
-{\cal M}(--){\cal M}^\ast(00)-{\cal M}(+0){\cal M}^\ast(0-)
+{\cal M}(-0){\cal M}^\ast(0+)],\cr
M^{Im}_{TL\overline{T}L}&=&\frac{1}{4}{\it Im}[{\cal M}(++){\cal
 M}^\ast(00)
+{\cal M}(--){\cal M}^\ast(00)+{\cal M}(+0){\cal M}^\ast(0-)
+{\cal M}(-0){\cal M}^\ast(0+)],\cr
M^{Im}_{\overline{T}LTL}&=&\frac{1}{4}{\it Im}[{\cal M}(++){\cal
 M}^\ast(00)
+{\cal M}(--){\cal M}^\ast(00)-{\cal M}(+0){\cal M}^\ast(0-)
-{\cal M}(-0){\cal M}^\ast(0+)],\cr
M^{Im}_{TLTL}&=&\frac{1}{4}{\it Im}[{\cal M}(++){\cal M}^\ast(00)
-{\cal M}(--){\cal M}^\ast(00)-{\cal M}(+0){\cal M}^\ast(0-)
+{\cal M}(-0){\cal M}^\ast(0+)],\cr
M^{Im}_{\overline{T}L\overline{T}L}&=&\frac{1}{4}{\it Im}[{\cal
 M}(++){\cal M}^\ast(00)
-{\cal M}(--){\cal M}^\ast(00)+{\cal M}(+0){\cal M}^\ast(0-)
-{\cal M}(-0){\cal M}^\ast(0+)],\cr
M^{Im}_{TTTT}&=&{\it Im}[{\cal M}(++){\cal M}^\ast(--)].
\label{amp2}
\end{eqnarray}

The squared amplitudes
\begin{equation}
M_{T\overline{T}}, M_{\overline{T}T},
       M_{\overline{T}L}, M_{L\overline{T}},
       M_{TL\overline{T}L}, M_{\overline{T}LTL},
       M^{Im}_{\overline{T}L\overline{T}L},
       M^{Im}_{TLTL}
\end{equation}
vanish if both the interaction responsible for the transition $V_1 V_2
\to W$ is parity conserving, i.e.\ if ${\cal M}(m,n)={\cal M}(-m,-n)$,
and a summation over the polarization of the final state $W$ is
performed. The luminosities $K^{Im}_{pol}$ vanish after integrating
over the azimuthal angle $\varphi_1$. We also note, that the squared
amplitudes $M^{Im}_{pol}$ are zero if all amplitudes ${\cal M}(m,n)$
can be chosen real. Therefore we restrict the following discussion to
the remaining eight luminosities
\begin{equation}
K_{TT},K_{\overline{TT}},K_{TL},K_{LT},K_{LL},K_{TLTL},
K_{\overline{T}L\overline{T}L},K_{TTTT}.
\end{equation}

The expression Eq.\ (\ref{terms19}) shows explicitly the trivial
factorization of the cross section into parts describing the
vector-boson emission from the incoming fermions and parts pertaining
to the vector-boson vector-boson scattering. These latter pieces,
combined with the phase space integral for the final state $W$, can be
interpreted as cross sections and correlations for virtual
vector-boson scattering processes:
\begin{equation}
\sigma_{pol}(q_1^2,q_2^2)=
             (2\pi)^4\frac{1}{2\kappa}\int d\rho_W M_{pol}
             \delta^{(4)}(q_1+q_2-p_W).
\label{sigvirt}
\end{equation}
In Eq.\ (\ref{sigvirt}) we included the 'flux factor' $1/2\kappa$ with
\begin{equation}
\kappa=\sqrt{{\cal W}^4 +q_1^4+q_2^4-2{\cal W}^2q_1^2-2{\cal W}^2q_2^2
-2q_1^2q_2^2},
\end{equation}
${\cal W}^4 \equiv ({\cal W}^2)^2$ and $q_j^4 \equiv (q_j^2)^2$, so
that (\ref{sigvirt}) leads to the correct expression for real
vector-boson scattering in the limit $q_j^2 \rightarrow M_j^2$.

In terms of the cross sections (\ref{sigvirt}) for virtual vector-boson
scattering, the cross section (\ref{sigma1}) for the two-fermion
initiated process is given by
\begin{eqnarray}
\sigma_{ff}&=&\left(\frac{\alpha}{2\pi}\right)^2\int\limits_{x_0}^1dx
\int\limits_{x}^1\frac{dz}{z}\int\limits_{-s(1-z)}^0\!\!\!\!\!\!\!dq_1^2
\int\limits_{-zs(1-\frac{x}{z})}^0\!\!\!\!\!\!dQ^2_2
\frac{1}{(q_1^2-M_1^2)^2}\frac{1}{(q_2^2-M_2^2)^2}\cr
&&\int\limits_0^{2\pi}\frac{d\varphi_1}{2\pi}\,\kappa
\sum\limits_{pol}c_{f,pol}\,K_{pol}\,\sigma_{pol}(q_1^2,q_2^2),
\label{sigma3}
\end{eqnarray}
where $\alpha$ is the fine structure constant.

Up to this point, the calculation has been exact without any
approximation. The basic assumption of the equivalent vector boson
method concerns the dependence of the off-shell cross sections
$\sigma_{pol}(q_1^2, q_2^2)$ on the off-shell masses $q_i^2$. For
transverse polarization it is certainly a good approximation to
identify $\sigma_{TT}(q_1^2, q_2^2)$ with its on-shell value
$\sigma_{TT}(M_1^2, M_2^2)$. However, for longitudinal polarizations,
$\sigma_{pol}(q_1^2, q_2^2)$ contains kinematic singularities at
$q_1^2 = 0$ and $q_2^2 = 0$, as can be seen from the explicit form
of the polarization vectors (Eq.\ (\ref{polC}) in Appendix \ref{app2}).
Therefore, for longitudinal polarization, the resulting factors
$M_i^2 / q_i^2$ should be taken into account explicitly. Apart from
this, there are good arguments from dispersion relation techniques
to believe that the extrapolation to off-shell masses is a smooth one.

We therefore make the assumption that the extrapolation to off-shell
masses can be described by simple proportionality factors
$f_{pol}(q_1^2,q_2^2)$ with $f_{pol}(M_1^2, M_2^2) = 1$. Taking also
the $q_i^2$-dependence of the flux factor $\kappa$ into account, we
write
\begin{equation}
\kappa\,\sigma_{pol}(q_1^2,q_2^2)=\tilde{\kappa}_0\,f_{pol}(q_1^2,
q_2^2)\,\sigma_{pol}(M_1^2,M_2^2),
\label{fvirt}
\end{equation}
where $\tilde{\kappa}_0$ is a flux factor for on-shell vector-boson
scattering to be specified below and $\sigma_{pol}(M_1^2,M_2^2)$ are
the cross-sections for on-shell vector-boson scattering evaluated at
the rescaled energy squared $(q_1+q_2)^2=xs$ of the vector-boson
vector-boson scattering process.

To describe the $q_j^2$-dependence of the off-shell cross sections,
we will consider the following specific forms of the proportionality
factors $f_{pol}$ which take into account the $q_j^2$-dependence of
the longitudinal polarization vectors $\epsilon_j(0)$:
\begin{eqnarray}
f_{TT}=f_{\overline{TT}}=f_{TTTT}=1,\cr
f_{TL}=\tfrac{M_2^2}{-q_2^2},\cr
f_{LT}=\tfrac{M_1^2}{-q_1^2},\cr
f_{LL}=\tfrac{M_1^2}{-q_1^2}\tfrac{M_2^2}{-q_2^2},\cr
f_{TLTL}=f_{\overline{T}L\overline{T}L}=
\tfrac{M_1}{\sqrt{-q_1^2}}\tfrac{M_2}{\sqrt{-q_2^2}}.
\label{fpol1}
\end{eqnarray}

We now introduce luminosities ${\cal L}_{pol}(x)$ which are
differential in the variable $x$, writing the differential cross
section in the form
\begin{equation}
\frac{d\sigma_{ff}}{dx}=
      \sum_{pol}{\cal L}_{pol}(x)\,\sigma_{pol}(M_1^2,M_2^2),
\end{equation}
with the luminosities
\begin{eqnarray}
{\cal L}_{pol}(x)&=&
         \left(\frac{\alpha}{2\pi}\right)^2\frac{\tilde{\kappa}_0}
{s}c_{f,pol}\,\int\limits_x^1\frac{dz}{z}\int\limits_{-s(1-z)}^0
\!\!\!\!\!\!\!dq_1^2
\int\limits_{-sz(1-\frac{x}{z})}^0\!\!\!\!\!\!dQ^2_2
\frac{1}{(q_1^2-M_1^2)^2}\frac{1}{(q_2^2-M_1^2)^2}\cr
&& \int\limits_0^{2\pi}\frac{d\varphi_1}{2\pi} \,
   f_{pol}(q_1^2,q_2^2)\,K_{pol}.
\label{lumis}
\end{eqnarray}
The luminosities ${\cal L}_{pol}(x)$ depend on $x$ and, since they
are dimensionless, on the masses of the vector bosons via the ratios
$M_1^2/s$ and $M_2^2/s$. As well, we have included the coupling
constants of the vector bosons in the definition. ${\cal L}_{pol}(x)
\,dx$ can be interpreted as the probability that the vector-boson pair
$V_1,$ $V_2$ with the specified polarization and with center-of-mass
energy in the interval $[xs, (x+dx)s]$ will be emitted from the
fermion pair $1$ and $2$.


\section{The exact Luminosities}

We evaluate the expressions (\ref{lumis}) adopting the forms
(\ref{fpol1}) for the behavior of the virtual cross-sections. No other
assumptions are made. The flux factor $\tilde{\kappa}_0$ is evaluated
at $q_j^2=M_j^2,j=1,2$, so that we have
\begin{equation}
\tilde{\kappa}_0=\kappa_0=\sqrt{{\cal W}^4+M_1^4+M_2^4-2{\cal W}^2M_1^2
-2{\cal W}^2M_2^2 -2M_1^2M_2^2}.
\end{equation}
We rewrite the phase space integral in (\ref{lumis}) in the following
way,
\begin{equation}
\int\limits_x^1\frac{dz}{z}\int\limits_{-s(1-z)}^0\!\!\!\!\!\!dq_1^2
\int\limits_{-sz(1-\frac{x}{z})}^0\!\!\!\!\!\!\!dQ^2_2
\int\limits_0^{2\pi}\frac{d\varphi_1}{2\pi}
=\int\limits_{-s+{\cal W}_1^2}^0 dq_1^2
 \int\limits_{-s+{\cal W}_2^2}^0 dq_2^2
 \int\limits_{\hat{x}s}^s\frac{d\mu_X}{\mu_X}
\int\limits_0^{2\pi}\frac{d\varphi_1}{2\pi},
\label{ps5}
\end{equation}
where we have introduced the variables $\hat{x} = \tfrac{\nu+K{\cal
W}}{s}$, with $\nu=q_1\cdot q_2=\frac{1}{2}({\cal W}^2-q_1^2-q_2^2)$,
${\cal W}=\sqrt{{\cal W}^2}$ and $K=\tfrac{\kappa}{2{\cal W}}$,
$K$ being the magnitude of the three-momentum of the vector-bosons
$V_1,V_2$ in their center-of-mass frame, and $\mu_X=M_X^2-q_1^2$.
The integration limits for $q_1^2$ and $q_2^2$ in (\ref{ps5}),
following from $(q_1^2+s)(q_2^2+s)>{\cal W}^2s$ (with $q_1^2<0$ and
$q_2^2<0$), are written with the help of ${\cal W}_1^2={\cal W}^2$
and ${\cal W}_2^2={\cal W}^2\tfrac{s}{s+q_1^2}$. The luminosities
vanish for $x < \tfrac{(M_1+M_2)^2}{s}$.

Using Eq.\ (\ref{ps5}), the expressions (\ref{lumis}) for the
luminosities become
\begin{equation}
{\cal L}_{pol}(x)=\left(\frac{\alpha}{2\pi}\right)^2\frac{\kappa_0}{s}
c_{f,pol}
\int\limits_{-s+{\cal W}_1^2}^0\!\!\!\!\!\!dq_1^2
\int\limits_{-s+{\cal W}_2^2}^0\!\!\!\!\!\!dq_2^2
\tfrac{q_1^2}{(q_1^2-M_1^2)^2}\tfrac{q_2^2}{(q_2^2-M_2^2)^2}\,
f_{pol}(q_1^2,q_2^2)\,J_{pol},
\label{lumis2}
\end{equation}
with the triple-differential luminosities---they are functions of
$x,q_1^2$ and $q_2^2$---
\begin{equation}
J_{pol}=
\tfrac{1}{q_1^2q_2^2}\int\limits_{\hat{x}s}^s\frac{d\mu_X}{\mu_X}
\int\limits_0^{2\pi}\frac{d\varphi_1}{2\pi} \,K_{pol},
\label{Jdef}
\end{equation}
and $K_{pol}$ were defined in (\ref{Kdef}). The integrations over $z$
and $\varphi_1$ in (\ref{Jdef}) can be performed analytically and the
results are given in (\ref{Jpols}).
We will discuss later which limiting
cases will lead to results already obtained in the literature.

The singularities of the integrands in Eq.\ (\ref{lumis2}) at $q_j^2
= M_j^2$ lead, after integration, to mass singular terms. In the
high-energy limit $s \gg M_j^2$ they appear either as familiar
logarithms $\ln (s/M_j^2)$, or as a pole singularity $1/M_j^2$. The
latter happens, e.g., in both masses for the $LL$-term, or in one of
the masses for the $TL$ and $LT$-luminosities. Since we will evaluate
the two-dimensional integration over $q_1^2$ and $q_2^2$ in
(\ref{lumis2}) numerically, specific care has to be taken of these
singularities. This is done by introducing new integration variables
$x_j$, $y_j$, and $z_j$ ($j=1,2$) depending on the type of the
singularity. The new variables are chosen such that the integration
region becomes the unit cube in two dimensions. Their relations to
$q_j^2$ are given by:
\begin{eqnarray}
q_j^2&=&M_j^2\left[1-\left(\tfrac{M_j^2+s-{\cal W}_j^2}{M_j^2}\right)^{
\textstyle x_j}\right] \nonumber \\
&=&M_j^2\left[1-\tfrac{M_j^2+s-{\cal W}_j^2}{(s-{\cal
 W}_j^2)(1-y_j)+M_j^2}\right] \nonumber \\
&=&(-s+{\cal W}_j^2)\,z_j,    ~~~~~~~  j=1,2\,\,,
\end{eqnarray}
and the luminosities (\ref{lumis2}) take the final form
\begin{eqnarray}
{\cal L}_{TT}(x)&=&
\left(\frac{\alpha}{2\pi}\right)^2(v_1^2+a_1^2)(v_2^2+a_2^2)
\frac{\kappa_0}{s}
\ln\left(\frac{M_1^2+s-{\cal W}_1^2}{M_1^2}\right)\int\limits_0^1dx_1
\int\limits_0^1dx_2\cr
&&\ln\left(\frac{M_2^2+s-{\cal W}_2^2}{M_2^2}\right)
\frac{q_1^2}{q_1^2-M_1^2}\frac{q_2^2}{q_2^2-M_2^2}J_{TT},
\nonumber \\
{\cal L}_{\overline{TT}}(x)&=&
\left(\frac{\alpha}{2\pi}\right)^2(2v_1a_1)(2v_2a_2)
\frac{\kappa_0}{s}
\ln\left(\frac{M_1^2+s-{\cal W}_1^2}{M_1^2}\right)\int\limits_0^1dx_1
\int\limits_0^1dx_2\cr
&&\ln\left(\frac{M_2^2+s-{\cal W}_2^2}{M_2^2}\right)
\frac{q_1^2}{q_1^2-M_1^2}\frac{q_2^2}{q_2^2-M_2^2}J_{\overline{TT}},
\nonumber \\
{\cal L}_{TL}(x)&=&\left(\frac{\alpha}{2\pi}\right)^2(v_1^2+a_1^2)
(v_2^2+a_2^2)
\frac{\kappa_0}{s}
\ln\left(\frac{M_1^2+s-{\cal W}_1^2}{M_1^2}\right)\int\limits_0^1dx_1
\int\limits_0^1dy_2\cr
&&\left(1-\frac{M_2^2}{M_2^2+s-{\cal W}_2^2}\right)
\frac{q_1^2}{q_1^2-M_1^2}J_{TL},
\nonumber \\
{\cal L}_{LT}(x)&=&\left(\frac{\alpha}{2\pi}\right)^2(v_1^2+a_1^2)
(v_2^2+a_2^2)\frac{\kappa_0}{s}
\left(1-\frac{M_1^2}{M_1^2+s-{\cal W}_1^2}\right)
\int\limits_0^1dy_1\int\limits_0^1dx_2\cr
&&\ln\left(\frac{M_2^2+s-{\cal W}_2^2}{M_2^2}\right)
\frac{q_2^2}{q_2^2-M_2^2}J_{LT},
\nonumber \\
{\cal L}_{LL}(x)&=&\left(\frac{\alpha}{2\pi}\right)^2(v_1^2+a_1^2)
(v_2^2+a_2^2)\frac{\kappa_0}{s}
\left(1-\frac{M_1^2}{M_1^2+s-{\cal W}_1^2}\right)
\int\limits_0^1dy_1\int\limits_0^1dy_2\cr
&&\left(1-\frac{M_2^2}{M_2^2+s-{\cal W}_2^2}\right)J_{LL},
\nonumber \\
{\cal L}_{TLTL}(x)&=&
\left(\frac{\alpha}{2\pi}\right)^2(v_1^2+a_1^2)(v_2^2+a_2^2)
\frac{\kappa_0}{s}\,M_1\,M_2\,
\ln\left(\frac{M_1^2+s-{\cal W}_1^2}{M_1^2}\right)\int\limits_0^1dx_1
\int\limits_0^1dx_2\cr
&&\ln\left(\frac{M_2^2+s-{\cal W}_2^2}{M_2^2}\right)
\frac{q_1^2}{q_1^2-M_1^2}\frac{q_2^2}{q_2^2-M_2^2}
\frac{J_{TLTL}}{\sqrt{-q_1^2}\sqrt{-q_2^2}},
\nonumber \\
{\cal L}_{\overline{T}L\overline{T}L}(x)&=&
\left(\frac{\alpha}{2\pi}\right)^2 (2v_1a_1)(2v_2a_2)
\frac{\kappa_0}{s}\,M_1\,M_2\,
\ln\left(\frac{M_1^2+s-{\cal W}_1^2}{M_1^2}\right)\int\limits_0^1dx_1
\int\limits_0^1dx_2\cr
&&\ln\left(\frac{M_2^2+s-{\cal W}_2^2}{M_2^2}\right)
\frac{q_1^2}{q_1^2-M_1^2}\frac{q_2^2}{q_2^2-M_2^2}
\frac{J_{\overline{T}L\overline{T}L}}{\sqrt{-q_1^2}\sqrt{-q_2^2}},
\nonumber \\
{\cal L}_{TTTT}(x)&=&\left(\frac{\alpha}{2\pi}\right)^2(v_1^2+a_1^2)
(v_2^2+a_2^2)\frac{\kappa_0}{s}
(s-{\cal W}_1^2)
\int\limits_0^1dz_1\int\limits_0^1dz_2\cr
&&(s-{\cal W}_2^2)
\frac{q_1^4}{(q_1^2-M_1^2)^2}\frac{q_2^4}{(q_2^2-M_2^2)^2}
\frac{J_{TTTT}}{q_1^2\,q_2^2},
\label{lumis3}
\end{eqnarray}
and the $J_{pol}$ are given by
\begin{eqnarray}
J_{TT}&=&\frac{8}{\kappa^4}\left[(2\nu^2(s+\nu)^2+q_1^2q_2^2(s^2+8s\nu
+q_1^2q_2^2))
\ln\left(\frac{1}{\hat{x}}\right)-6s^2\nu^2-4s\nu^3+2\nu^4\right.\cr
&&\left.+q_1^2q_2^2(-3s^2+4s\nu+6\nu^2+q_1^2q_2^2)
+K{\cal W}(3s^2\nu+8s\nu^2+2\nu^3+q_1^2q_2^2(4s+\nu))\right],\cr
J_{\overline{TT}}&=&\frac{4}{\kappa^2}\left[(2s\nu+\nu^2+q_1^2q_2^2)
\ln\left(\frac{1}{\hat{x}}\right)-4s\nu+2\nu^2+2q_1^2q_2^2
+2K{\cal W}(s+\nu)\right],\cr
J_{TL}=J_{LT}&=&
\frac{4}{\kappa^4}\left[(4s^2\nu^2+8s\nu^3+2q_1^2q_2^2(s^2+8s\nu
+3\nu^2))\ln\left(\frac{1}{\hat{x}}\right)-13s^2\nu^2-4s\nu^3
\right.\cr
&&\left.
+2\nu^4 +q_1^2q_2^2(-5s^2+4s\nu+13\nu^2+3q_1^2q_2^2)\right.\cr
&&\left.
+2K{\cal W}(3\nu^2+8s\nu^2+\nu^3+2q_1^2q_2^2(2s+\nu))\right],\cr
J_{LL}&=&\frac{8}{\kappa^4}\left[(2s^2\nu^2+4s\nu^3+q_1^2q_2^2(s^2+8s\nu
+2\nu^2+q_1^2q_2^2))\ln\left(\frac{1}{\hat{x}}\right)
-7s^2\nu^2\right.\cr&&\left.+q_1^2q_2^2(-2s^2+7\nu^2+2q_1^2q_2^2)
+K{\cal W}(3s^2\nu+8s\nu^2+q_1^2q_2^2(4s+3\nu))\right],\cr
J_{TLTL}&=&\frac{32}{\kappa^4}\sqrt{-q_1^2}\sqrt{-q_2^2}
\left[(3s^2\nu+9s\nu^2+\nu^3+q_1^2q_2^2(3s+2\nu))
\ln\left(\frac{1}{\hat{x}}\right)-\frac{\nu^2(s+\nu)}{\hat{x}}\right.\cr
&&\left.-8s^2\nu+3\nu^3+q_1^2q_2^2(s+6\nu)
+K{\cal W}(2s^2+11s\nu+2\nu^2+q_1^2q_2^2)\right],\cr
J_{\overline{T}L\overline{T}L}&=&
\frac{8}{\kappa^2}\sqrt{-q_1^2}\sqrt{-q_2^2}
\left[(s+\nu)\ln\left(\frac{1}{\hat{x}}\right)-\frac{\nu}{\hat{x}}
+2\nu-s+K{\cal W}\right],\cr
J_{TTTT}&=&\frac{4}{\kappa^4}q_1^2q_2^2
\left[(3s^2+12s\nu+2\nu^2+q_1^2q_2^2)
\ln\left(\frac{1}{\hat{x}}\right)+\frac{\nu^2}{\hat{x}^2}
-\frac{5s\nu+4\nu^2}{\hat{x}}\right.\cr
&&\left.-5s^2+4s\nu+6\nu^2+3q_1^2q_2^2+K{\cal W}(8s+3\nu)\right].
\label{Jpols}
\end{eqnarray}
For the case of two-photon processes initiated by electron-electron
scattering, analogous expressions have been derived in \cite{bonneau}.
Our results are related to the corresponding $\hat{J}_{pol}$ from
\cite{bonneau} by $J_{pol} = \frac{1}{x^2} \hat{J}_{pol}$ for
$pol=TT$, $TL$ and $LL$, $J_{TTTT} = \frac{1}{x^2} \hat{J}_{TT}^{ex}$
and $J_{TLTL}=\frac{2}{x^2}\hat{J}_{LT}^{ex}$ (note that we have
neglected the fermion masses). We finally remark that, for $M_1=M_2$,
we have ${\cal L}_{LT}(x)={\cal L}_{TL}(x)$.

The integrals in Eq.\ (\ref{lumis3}) are well-suited for numerical
evaluation. Their integrands contain no singularities; instead, the
poles of order one show up as logarithms of the form $\ln\left((M_j^2 +
s - {\cal W}_j^2) / M_j^2 \right)$, while the poles of order two (which
would by themselves lead to a factor $M_j^{-2}$) have been canceled by
corresponding factors $M_j^2$ included in our assumptions for the
behaviour of the $f_{pol}(q_1^2,q_2^2)$, Eq.\ (\ref{fpol1}). Since the
expressions Eq.\ (\ref{lumis3}) involve two-dimensional numerical
integrations of the momentum transfers $q_1^2$ and $q_2^2$, it would be
straightforward to replace the model assumptions Eq.\ (\ref{fpol1}) by
better ones if required. The contribution from the leading
singularities would not change then; however, subleading terms
(non-logarithmic contributions for transverse polarization,
logarithmic contributions for longitudinal polarization) are
model-dependent. For the cases of Higgs production and heavy quark
production, modifications of single-$W$ boson distributions following
from the exact off-shell behaviour of the corresponding hard cross
sections have been studied in \cite{cortese}.


\section{Convolutions of single-vector-boson distributions}

Since helicities of massive particles are not Lorentz-invariant, the
polarization vectors have to be defined in a definite reference frame,
which we chose to be the center-of-mass system of the two vector bosons.
Therefore, the ${\cal C}_i$ and ${\cal S}_i$ depend on both momentum
transfers $q_1^2$ and $q_2^2$ at the same time. This means that the
emission of a vector boson $V_1$ with definite helicity from fermion
1 is not independent from the off-shell mass of the second vector boson
$V_2$, and the two-boson luminosities do not factorize into single-boson
densities. However, since at high energies the process is dominated
by small momentum transfers, it seems justified to neglect this mutual
dependence on $q_i^2$. Then the expressions (\ref{lumis}) for the
two-vector-boson luminosities reduce to convolutions of
single-vector-boson densities. These single-vector-boson distributions
have been reported in \cite{tung1}.

To be specific, we consider the following simplifications:
\begin{enumerate}
\item Set $q_2^2=0$ in ${\cal C}_1(m,m')$ and
${\cal S}_1(m,m')$;\label{sim1}
\item Set $q_1^2=0$ in ${\cal C}_2(n,n')$ and
${\cal S}_2(n,n')$;\label{sim2}
\item Set $Q^2_2=q_2^2$ in Eq.\ (\ref{lumis}), i.e.\ omit the
factor $(1 - q_1^2/M_x^2)^{-1}$ in the definition Eq.\ (\ref{bigq2})
of $Q^2_2$.
\end{enumerate}
In addition, we evaluate the flux factor $\tilde{\kappa}_0$ at $q_1^2=0$
and $q_2^2=0$, i.e.\ we choose $\tilde{\kappa}_0={\cal W}^2$. Note that
with the simplifications \ref{sim1} and \ref{sim2}, the luminosities
for the non-diagonal squared amplitudes, ${\cal L}_{TLTL}(x),
{\cal L}_{\overline{T}L\overline{T}L}(x)$ and ${\cal L}_{TTTT}(x)$
vanish.

With these simplifications the integrals over $q_1^2$ and $Q^2_2$ in
(\ref{lumis}) can be carried out independently and the luminosities
(\ref{lumis}) take the factorized form
\begin{equation}
{\cal L}_{kl}(x)=\int\limits_x^1\frac{dz}{z}\,P^1_k(z,M_1^2)\,
P^2_l\left(\frac{x}{z},\frac{M_2^2}{z}\right),
\label{conv}
\end{equation}
where $k, l = T$, $\overline{T}$, $L$ and the functions $P^i_T$,
$P^i_{\overline{T}}$ and $P^i_L$ are the single-vector-boson
distributions of \cite{tung1}, explicit forms of which are
\begin{eqnarray}
P^j_T(z,M^2)&=&\frac{\alpha}{2\pi}(v_j^2+a_j^2)\tfrac{z}{2}
\int\limits_{-s(1-z)}^0\tfrac{d(q^2)\,(-q^2)(c_0^2+1)}{(q^2-M^2)^2},\cr
P^j_{\overline{T}}(z,M^2)&=&\tfrac{\alpha}{2\pi}(2v_ja_j)\,z
\int\limits_{-s(1-z)}^0\tfrac{d(q^2)\,(-q^2)\,c_0}{(q^2-M^2)^2},\cr
P^j_L(z,M^2)&=&\frac{\alpha}{2\pi}(v_j^2+a_j^2)\,M^2\tfrac{z}{2}
\int\limits_{-s(1-z)}^0\tfrac{d(q^2)\,s_0^2}{(q^2-M^2)^2},
\label{distri}
\end{eqnarray}
with
\begin{equation}
c_0 = \tfrac{2-z+\tfrac{q^2}{s}}{z-\tfrac{q^2}{s}}
\quad\quad \mbox{and} \quad\quad
s_0 = 2\tfrac{\sqrt{1-z+\tfrac{q^2}{s}}}{z-\tfrac{q^2}{s}}.
\end{equation}
The integrals in (\ref{distri}) can be performed analytically and the
results have been given in \cite{tung1}\footnote{Also the distributions
    of \cite{lindfors} are exact for processes with only one internal
    vector boson which couples to the amplitude for the hard scattering
    subprocess like a fermion. This specific assumption in
    \cite{lindfors} is the only difference between the distributions
    of \cite{lindfors} and \cite{tung1}.}. The quantities $P^j_k(z,M^2)$
are the probability densities for the emission of a vector boson with
mass $M$ from a fermion $j$ with couplings $v_j$ and $a_j$. The scaling
variable $z$ describes the invariant mass squared remaining after the
emission of the vector boson $V_1$ from fermion $1$. Since $q_2^2$ has
been neglected in describing the emission, the center-of-mass system
$C$ of the two vector bosons is related to the center-of-mass system of
vector boson $V_1$ and fermion $2$ by a boost in the direction of the
fermion $2$. Therefore, the helicities of the vector boson $V_1$,
originally defined in the center-of-mass system $C$, agree in the two
reference systems. The same line of thought applies to the emission of
vector boson $V_2$ from fermion $2$ with the scaling variable $z$ being
replaced by $x/z$.

In summary, the luminosities (\ref{lumis}) can be written as
convolutions (\ref{conv}) of single-vector-boson distributions
(\ref{distri}) if one neglects the mutual effects of the variation
of the off-shellness of one of the vector bosons on the probability for
the emission of the other vector boson. The luminosities for the
off-diagonal squared amplitudes vanish in this case.


\section{Leading Logarithmic Approximation}

Further approximations in Eqs.\ (\ref{lumis3}) allow to derive
simplified expressions which have often been used in the literature and
are referred to as the leading logarithmic approximation (LLA). The
approximation consists in neglecting the off-shell masses $q_i^2$ in
$J_{pol}$ and performing a high-energy limit, $s \gg M_j^2$. To be
precise, with the following substitutions in (\ref{lumis3}),
\begin{eqnarray}
\ln\left(\frac{M_j^2+s-{\cal W}_j^2}{M_j^2}\right)&\to&
\ln\left(\frac{s}{M_j^2}\right),\cr
\left(1-\frac{M_j^2}{M_j^2+s-{\cal W}_j^2}\right)&\to&1,\cr
\frac{q_j^2}{q_j^2-M_j^2}&\to&1,\cr
\kappa_0&\to&{\cal W}^2,
\end{eqnarray}
one obtains
\begin{eqnarray}
{\cal L}_{TT}(x)&\to&\left(\frac{\alpha}{2\pi}\right)^2(v_1^2+a_1^2)
(v_2^2+a_2^2)\frac{1}{x}\left[(2+x)^2
\ln\left(\frac{1}{x}\right)
-2(1-x)(3+x)\right] \cdot \cr
&&\ln\left(\frac{s}{M_1^2}\right) \ln\left(\frac{s}{M_2^2}\right),\cr
{\cal L}_{\overline{TT}}(x)&\to&
\left(\frac{\alpha}{2\pi}\right)^2(2v_1a_1)
(2v_2a_2)\left[(4+x)\ln\left(\frac{1}{x}\right)
-4(1-x)\right]\ln\left(\frac{s}{M_1^2}\right)\ln\left(\frac{s}{M_2^2}
\right),\cr
{\cal L}_{TL}(x)&\to&\left(\frac{\alpha}{2\pi}\right)^2(v_1^2+a_1^2)
(v_2^2+a_2^2)\frac{1}{x}\left[4(1+x)\ln\left(\frac{1}{x}\right)
-(1-x)(7+x)\right]\ln\left(\frac{s}{M_1^2}\right),\cr
{\cal L}_{LT}(x)&\to&\left(\frac{\alpha}{2\pi}\right)^2(v_1^2+a_1^2)
(v_2^2+a_2^2)\frac{1}{x}\left[4(1+x)\ln\left(\frac{1}{x}\right)
-(1-x)(7+x)\right]\ln\left(\frac{s}{M_2^2}\right),\cr
{\cal L}_{LL}(x)&\to&\left(\frac{\alpha}{2\pi}\right)^2(v_1^2+a_1^2)
(v_2^2+a_2^2)\frac{4}{x}\left[(1+x)\ln\left(\frac{1}{x}\right)
-2(1-x)\right].
\label{lelo}
\end{eqnarray}
Expressions for ${\cal L}_{TT}$, ${\cal L}_{TL}$ and ${\cal L}_{LL}$
have been given already in \cite{lindfors3} and the complete set of
luminosities including ${\cal L}_{\overline{T}T}$, ${\cal L}_{
\overline{T}L}$ and ${\cal L}_{L\overline{T}}$ can be found in
\cite{renard}. In a similar way, LLA expressions for
single-vector-boson distributions can be obtained from the exact ones,
Eq.\ (\ref{distri}). Their convolutions lead again to Eq.\ (\ref{lelo}).

These formulae are obtained from the exact ones by taking into account
only the contributions from the singularities at $q_j^2\to 0$ to the
$q_j^2$-integrals and neglecting the contribution from other regions in
the $q_1^2,q_2^2$ integration. The choice of $s$ in the arguments of
the logarithms is arguable; many other choices are also acceptable in
the leading logarithmic approximation and have been used in the
literature. For example, $xs$ as argument instead of $s$ has been
advocated in \cite{lindfors,lindfors3}, since the quantity $s-{\cal
W}_2^2$ varies in the whole interval $[0,s]$ as $q_1^2$ varies within
its limits. We have checked numerically that the LLA with this choice
deviates less from the exact calculation. The deviation for $x\to 1$
can be improved by choosing $x(1-x)s$ instead of $xs$ in the argument
of the logarithms. This choice is motivated by interpreting the
approximation as resulting from a zero-mass limit and noting that
$s-{\cal W}^2 = (1-x)s$. We will use this form in our numerical
examples.

Related to the different possible choices of the argument of the
logarithm is the interpretation of the scaling variable $x$. In
\cite{dawson,kane2,renard,godbole}, the scaling variable $x$ was
defined as the ratio of the vector-boson energy and the energy of the
fermion from which it is emitted. With this definition, the relation
$\hat{s}=xs$ between the fermion scattering energy and the subprocess
energy only holds strictly if the vector-boson is emitted in the
forward direction. These versions thus imply a small angle
approximation. In addition, the mentioned distributions differ by
various additional approximations. The distributions of \cite{kane2}
neglect terms of the order ${\cal O} \left(M^2_i/s\right)$. In
\cite{dawson,godbole}\footnote{The distributions of \cite{godbole}
            supplement those of \cite{dawson} by the distribution
            function $P_{\overline{T}}$ (see Eq.\ (\ref{distri})).},
the calculation was performed using a longitudinal polarization vector
for on-shell vector bosons, whereas in \cite{renard}
$\epsilon^{\mu}(0)$ was defined taking into account that the vector
bosons have off-shell masses $-q_j^2$. This and a more sophisticated
assumption concerning the off-shell behaviour of the hard scattering
cross section in \cite{renard} is the reason for the difference
between the distribution functions for longitudinal polarization in
\cite{dawson,godbole} and \cite{renard}. The distribution for
transversely polarized vector bosons in \cite{dawson,godbole} and
\cite{renard} agree with each other (after correcting misprints in the
latter reference). Of course, all distributions agree in the leading
logarithmic approximation.


\section{Numerical Results}

In presenting numerical results for luminosities of vector-boson pairs,
we restrict ourselves to the representative case of $e^+e^-$
annihilation. In our examples for the numerical evaluation we used
$\alpha=1/137$, $M_W=80.2$ GeV, $M_Z=91.2$ GeV and the fermion
vector-boson couplings are determined using the weak mixing angle as
given by $\cos\theta_W = M_W/M_Z$. In Figs.\ 2 and 3 we show the exact
luminosities (\ref{lumis3}) for finding a $W^+W^-$ pair in an $e^+e^-$
pair of $\sqrt{s}=2$ TeV. The luminosity ${\cal L}_{TT}$ for
transversely polarized $W^{\pm}$ is the biggest one, followed by ${\cal
L}_{TL}$ and ${\cal L}_{LL}$. From Fig.\ 3 one concludes that the
non-diagonal luminosities ${\cal L}_{TLTL}$ and ${\cal L}_{TTTT}$
are comparable in size with the diagonal ones and thus can not be
neglected. The parity violating luminosity ${\cal L}_{\overline{TT}}$
varies comparatively little with $x$ at not too high $x$, and at higher
$x$ it becomes equal to the $TT$ luminosity.

In order to estimate the improvement obtained by using the exact
luminosities as compared to former simpler approaches, we show in the
following series of figures ratios of the exact results and the
convolutions of the exact single-vector-boson distributions from
\cite{tung1} as well as their LLA versions. The ratio of the
convolutions (\ref{conv}) and the exact luminosities is shown in Fig.
4 for a $W^+W^-$ pair in a $2$ TeV $e^+e^-$ pair. The discrepancy grows
with decreasing $x$ and is largest for transverse polarizations in
which case it reaches a factor of $2.4$ at $x=0.01$. At higher energies
the agreement between the two versions is better as seen in Fig.\ 5
where the same ratio is shown for a value $\sqrt{s}= 4$ TeV, which is a
typical $q\bar{q}$ sub-process energy in $pp$ collisions at $14$ TeV.
However, the ratio of the $TT$ luminosities for $x=0.01$ is still $1.6$
(this corresponds to the production of a final state $W$ of $400$ GeV).

Fig.\ 6 shows the ratio of the LLA version of the luminosities, Eq.\
(\ref{lelo}), and the exact formulae for $W^+W^-$ in $e^+e^-$ at $2$
TeV. The LLA versions always overestimate the exact results by far and
only for the $LL$ luminosity at not too small values of $x$ the LLA
might be useful. We note that the disagreement at $x\rightarrow 1$
would have been larger if we had used $xs$ instead of $x(1-x)s$ in the
argument of the logarithms.

We also present some results relevant for a $500$ GeV $e^+e^-$ collider.
Figs.\ 7 and 8 show the luminosities for a $W^+W^-$ pair as a function
of the $W^+W^-$ pair invariant mass ${\cal W}$ related to $x$ by
${\cal W}^2=xs$. The luminosities reach their highest value not far from
threshold. The behavior of the different polarizations with varying
$x$ is as described for the $2$ TeV case. There is a resemblance
between the pairs $TT$, $\overline{TT}$ and $TLTL$,
$\overline{T}L\overline{T}L$. In both cases, the luminosity
proportional to the product of vector and axial-vector coupling is
smaller than its partner at low $x$ but then joins it at high $x$.
Finally, Figs.\ 9 and 10 show the luminosities for a $ZZ$ pair. The
major changes as compared to the $W^+W^-$ case are due to the change in
the vector-boson couplings, while the changes due to the different
vector-boson masses are small. The $ZZ$ luminosities are more than an
order of magnitude smaller than the $W^+W^-$ luminosities. Owing to the
small vector coupling of the $Z$, the luminosities which are
proportional to the product of vector and axial-vector coupling are
negligible.

In summary, only the luminosities for longitudinally polarized vector
boson pairs in regions of high $\sqrt{s}$ and $x$ might be described
by the convolutions or the LLA. For luminosities involving transverse
polarizations, neither of these two approximations reproduces the
exact calculations with a reasonable accuracy. The disagreement becomes
worse with decreasing $x$ and decreasing $\sqrt{s}$.

To obtain luminosities relevant for deep-inelastic lepton nucleon
scattering or for processes at hadron colliders, one would have to
adjust the factors in (\ref{lumis3}) containing the vector and
axial-vector coupling constants and, in addition, to fold the
luminosities with quark distribution functions. This would result
in luminosities for vector-boson pairs in an $ep$, $pp$ or $p\bar{p}$
initial state.


\section{Conclusion}

We have derived exact distribution functions for a pair of vector
bosons inside a pair of fermions. In contrast to previously used
approximations, our distributions take into account the mutual
influence of the emission of one boson on the emission of the other.
The commonly used leading logarithmic approximation and a convolution
of exact distribution functions for single vector bosons inside
fermions are obtained if one neglects regions in phase space in which
the virtual vector bosons have four-momenta squared much larger than
their squared masses. We have shown that for transverse polarizations
of the vector bosons, these approximations do not reproduce the exact
calculation with a reasonable accuracy.

Our results are obtained from an exact calculation of a subset of
Feynman diagrams without the need to introduce any approximation except
specific assumptions for the off-shell behaviour of vector-boson
scattering cross sections. A different off-shell behaviour could be
taken into account in our formalism without additional complications.
                           Of course, in order to obtain complete
predictions for cross sections of vector-boson production in $e^+e^-$
or hadron colliders, one would have to add contributions from Feynman
diagrams which are not of the type as shown in Fig.\ 1, as for example
$q\bar{q}$ annihilation or bremsstrahlung processes. These additional
contributions might become particularly important at smaller energies.

Finally one should note that we did not attempt to take into account
any kind of experimental cuts on kinematical variables for final state
particles, like transverse momenta or rapidities. These cuts would,
first of all, enter in the expressions for the vector-boson scattering
cross sections. As far as experimental cuts on final state momenta
imply restrictions also for the momentum transfers $q_i^2$, or
the scale variable $x$, it would be straightforward to modify our
expressions for the luminosities accordingly.


\begin{appendix}

\section{Breit-Systems and Polarization Vectors}

\subsection{Definition of Reference Frames\label{app1}}

The four-momenta in the center-of-mass system $C$ of $V_1$ and $V_2$
are
\begin{equation}
\left(q_1^{C}\right)^{\mu}=(k_0;0,0,K), ~~~~
\left(q_2^{C}\right)^{\mu}=(q_0;0,0,-K),
\end{equation}
with $k_0 = ({\cal W}^2+q_1^2-q_2^2)/ 2{\cal W}$ and $q_0 =
({\cal W}^2-q_1^2+q_2^2)/ 2{\cal W}$. For simplicity, we assume
that the final state $W$ produced via the 2-boson process allows to
specify the $x$- and $y$-axes of a coordinate system. If the state $W$
decays into $n$ particles with momenta $k_i$, we choose this system
such that the $y$-component of one specific four-momentum, say $k_s$,
of the set of $k_i$ vanishes and its $x$-component is non-negative.

We define two Breit systems, a system $B_1$ in which $q_1$ has only
a non-vanishing $z$-component and $\vec{l_2}$ points in the negative
$z$-direction, and a system $B_2$ in which $q_2$ has only a non-zero
$z$-component and $\vec{q_1}$ points in the negative $z$-direction. The
four-momenta in $B_1$ are
\begin{eqnarray}
\left(l_1^{B_1}\right)^{\mu}&=&\frac{\sqrt{-q_1^2}}{2}
           (c_h;-s_h\,\cos\varphi_1,-s_h\,\sin\varphi_1,1),\cr
\left(p_1^{B_1}\right)^{\mu}&=&\frac{\sqrt{-q_1^2}}{2}
           (c_h;-s_h\,\cos\varphi_1,-s_h\,\sin\varphi_1,-1),\cr
\left(q_1^{B_1}\right)^{\mu}&=&(0;0,0,\sqrt{-q_1^2}),\cr
\left(l_2^{B_1}\right)^{\mu}&=&
           \frac{\mu_X}{2\sqrt{-q_1^2}}(1;0,0,-1),\cr
{p'}^{\mu} \equiv p_W^{\mu} + p_2^{\mu}&=&
    \frac{1}{2\sqrt{-q_1^2}}(\mu_X;0,0,-M_X^2-q_1^2),
\label{Breit1}\end{eqnarray}
with $c_h=\tfrac{2s}{\mu_X}-1$, $s_h=\sqrt{c_h^2-1} = 2\tfrac{\sqrt{s}}
{\mu_X} \sqrt{s-\mu_X}$, and $\mu_X = M_X^2 - q_1^2$. The overall
azimuth of the system is defined by choosing the $y$-component of
$q_2^{B_1}$ to be zero and its $x$-component non-negative, so that
\begin{equation}
\left(q_2^{B_1}\right)^{\mu}=
\left(q_0'; \frac{\sqrt{-q_2^2}\beta}{\mu_X}, 0,
-\frac{\nu}{\sqrt{-q_1^2}}\right),
\end{equation}
with
$q_0'=\tfrac{1}{\sqrt{-q_1^2}}\left(\nu -
\tfrac{q_1^2q_2^2}{\mu_X}\right)$ and
$\beta=\sqrt{\mu_X^2-2\nu\mu_X+q_1^2q_2^2}$.

The four-momenta in $B_2$ are
\begin{eqnarray}
\left(l_2^{B_2}\right)^{\mu}&=&
\frac{\sqrt{-q_2^2}}{2}(c_h';-s_h'\,\cos\varphi_2,-s_h'
\,\sin\varphi_2,1),\cr
\left(p_2^{B_2}\right)^{\mu}&=&
\frac{\sqrt{-q_2^2}}{2}(c_h';-s_h'\,\cos\varphi_2,-s_h'
\,\sin\varphi_2,-1),\cr
\left(q_2^{B_2}\right)^{\mu}&=&
\left(0;0,0,\sqrt{-q_2^2}\right),\cr
\left(q_1^{B_2}\right)^{\mu}&=&
\frac{1}{2\sqrt{-q_2^2}}(\kappa;0,0,-2\nu),\cr
\left(p_W^{B_2}\right)^{\mu}&=&
\frac{1}{2\sqrt{-q_2^2}}(\kappa;0,0,-2{\cal W}\,q_0),
\label{Breit2}\end{eqnarray}
with $c_h'=\tfrac{2}{\kappa}(\mu_X-\nu)$ and $s_h'=\sqrt{(c_h')^2-1}=
\tfrac{2\beta}{\kappa}$. The overall azimuth of the system $B_2$ is
defined by choosing the $y$-component of the same four-momentum $k_s$
as employed in defining the system $C$ equal to zero and its
$x$-component non-negative.

\subsection{Polarization Vectors\label{app2}}

The polarization vectors for the helicity eigenstates of the vector
bosons $V_j$ in the system $C$ using the Jacob and Wick phase
conventions are
\begin{eqnarray}
\left(\epsilon_1^C\right)^{\mu}(\pm) &=&
     \frac{1}{\sqrt{2}}(0;\mp 1,-i,0),\cr
\left(\epsilon_1^C\right)^{\mu}(0)   &=&
     \frac{1}{\sqrt{-q_1^2}}(K;0,0,k_0),\cr
\left(\epsilon_2^C\right)^{\mu}(\pm) &=&
     \frac{1}{\sqrt{2}}(0;\pm 1,-i,0),\cr
\left(\epsilon_2^C\right)^{\mu}(0)   &=&
     \frac{1}{\sqrt{-q_2^2}}(-K;0,0,q_0).
\label{polC}
\end{eqnarray}
By applying an appropriate coordinate transformation,
the polarization vectors for $V_1$ in the system $B_1$ are
found to be
\begin{eqnarray}
\left(\epsilon_1^{B_1}\right)^{\mu}(\pm) &=&
                          \frac{1}{\sqrt{2}}e^{\mp\,i\,\varphi_2}
\left(\mp\tilde{\sigma}s_y; \mp\frac{\sqrt{-q_1^2}q_0'}{K{\cal W}}, -i,
0\right),\cr
\left(\epsilon_1^{B_1}\right)^{\mu}(0) &=&
                        \frac{\sqrt{-q_1^2}}{K{\cal W}}
\left(q_0';\frac{\sqrt{-q_2^2}\beta}{\mu_X},0,0\right),
\label{polB1}
\end{eqnarray}
with $\tilde{\sigma}s_y = \sqrt{q_1^2 q_2^2} \beta / (\mu_X K {\cal
W})$. Likewise, the polarization vectors for $V_2$ in $B_2$ are found
to be
\begin{eqnarray}
\left(\epsilon_2^{B_2}\right)^{\mu}(\pm)&=&
\frac{1}{\sqrt{2}}(0;\pm 1,i,0),\cr
\left(\epsilon_2^{B_2}\right)^{\mu}(0)&=&
(-1;0,0,0).
\label{polB2}
\end{eqnarray}

\section{Five-Fold Differential Luminosities\label{appB}}

Here we give explicit expressions needed to determine the five-fold
differential luminosities $K_{pol}$ of Eq.\ (\ref{Kdef}). The helicity
tensors ${\cal C}_j(m,m')$ and ${\cal S}_j(m,m')$, defined in Eqs.\
(\ref{Cexp}) and (\ref{Sexp}), are evaluated most easily in their
respective Breit systems $B_j$ using the expressions (\ref{Breit1}) and
(\ref{Breit2}) for the four-momenta and the expressions (\ref{polB1})
and (\ref{polB2}) for the polarization vectors. The results are:
\begin{eqnarray}
{\cal C}_1(++) &=& -\frac{q_1^2}{4}
  \left[c_h^2+1+\frac{4q_1^2q_2^2\,\beta^2}
  {\mu_X^2\kappa^2}(c_h^2+s_h^2\cos^2\varphi_1)\right.\cr
&&\left.+\frac{8\,c_h\,s_h\sqrt{-q_1^2}\sqrt{-q_2^2}\,\beta}
              {\mu_X^2\kappa^2}
  (\nu\mu_X-q_1^2q_2^2)\cos\varphi_1\right],\cr
{\cal C}_1(00) &=& -\frac{q_1^2}{2}
  \left[s_h^2+\frac{4q_1^2q_2^2\,\beta^2}
  {\mu_X^2\kappa^2}(c_h^2+s_h^2\cos^2\varphi_1)\right.\cr
&&\left.-\frac{8\,c_h\,s_h\sqrt{-q_1^2}\sqrt{-q_2^2}\,\beta}
              {\mu_X^2\kappa^2}
  (\nu\mu_X-q_1^2q_2^2)\cos\varphi_1\right],\cr
{\cal C}_1(+-) &=& \frac{q_1^2}{2}
  \left[\frac{2q_1^2q_2^2\,\beta^2}{\mu_X^2
  \kappa^2}(c_h^2+s_h^2\cos^2\varphi_1)+s_h^2\left(\cos^2\varphi_1
  -\frac{1}{2}\right) \right.\cr
&&\left.+\frac{4\,c_h\,s_h\sqrt{-q_1^2}\sqrt{-q_2^2}\,\beta}
              {\mu_X^2\kappa^2}
  (\nu\mu_X-q_1^2q_2^2)\cos\varphi_1\right.\cr
&&\left.-2\,i\frac{s_h}{\mu_X\kappa}
  (c_h\sqrt{-q_1^2}\sqrt{-q_2^2}\,\beta
  +s_h\,(\nu\mu_X-q_1^2q_2^2)\cos\varphi_1)\sin\varphi_1\right],\cr
{\cal C}_1(+0) &=& \frac{q_1^2}{\sqrt{2}}
  \left[\frac{2\sqrt{-q_1^2}\sqrt{-q_2^2}
  \,\beta}{\mu_X^2\kappa^2}
  (\nu\mu_X-q_1^2q_2^2)(c_h^2+s_h^2\cos^2\varphi_1)\right.\cr
&&\left.+\frac{2\,c_h\,s_h}{\mu_X^2\kappa^2}
  ((\nu\mu_X-q_1^2q_2^2)^2+q_1^2q_2^2\,\beta^2)\cos\varphi_1\right.\cr
&&\left.-i\frac{s_h}{\mu_X\kappa}
  (c_h(\nu\mu_X-q_1^2q_2^2)+s_h\,\sqrt{-q_1^2}\sqrt{-q_2^2}\,
  \beta\cos\varphi_1) \sin\varphi_1\right],\cr
{\cal S}_1(++) &=& -\frac{q_1^2}{2}
  \left[2\,\frac{\nu\mu_X-q_1^2q_2^2}
  {\mu_X\kappa}\,c_h+2\,\frac{s_h\,\sqrt{-q_1^2}\sqrt{-q_2^2}\,\beta}
  {\mu_X\kappa} \cos\varphi_1\right],\cr
{\cal S}_1(+0) &=& \frac{q_1^2}{\sqrt{2}}\left[\frac{c_h\,\sqrt{-q_1^2}
  \sqrt{q_2^2}\,\beta}{\mu_X\kappa}
  +\frac{s_h}{\mu_x\kappa}(\nu\mu_X-q_1^2q_2^2)
  \cos\varphi_1-i\frac{s_h}{2}\sin\varphi_1\right];
\label{cs}
\end{eqnarray}
\begin{eqnarray}
{\cal C}_2(++)&=&-\frac{q_2^2}{4}\left((c_h')^2+1\right),\cr
{\cal C}_2(00)&=&-\frac{q_2^2}{2}(s_h')^2,\cr
{\cal C}_2(+-)&=&\frac{q_2^2}{4}(s_h')^2,\cr
{\cal C}_2(+0)&=&\frac{q_2^2}{2\sqrt{2}}\,c_h'\,s_h',\cr
{\cal S}_2(++)&=&-\frac{q_2^2}{2}c_h',\cr
{\cal S}_2(+0)&=&\frac{q_2^2}{2\sqrt{2}}s_h'.
\label{csprime}
\end{eqnarray}

\end{appendix}



\newpage

{\Large \bf Figure Caption}

\begin{list}{}{\setlength{\leftmargin}{10mm}%
\setlength{\labelwidth}{15mm}}
\renewcommand{\baselinestretch}{0.5}
\item[Fig.\ 2: \hfill]
Luminosities ${\cal L}_{TT}(x)$, ${\cal L}_{\overline{TT}}(x)$,
${\cal L}_{LT}(x)$, and ${\cal L}_{LL}(x)$ for a $W^+W^-$ pair in
$e^+e^-$ collisions at $\sqrt{s} = 2$ TeV.

\item[Fig.\ 3: \hfill]
Luminosities ${\cal L}_{TLTL}(x)$, ${\cal
L}_{\overline{T}L\overline{T}L}(x)$, and ${\cal L}_{TTTT}(x)$
for a $W^+W^-$ pair in $e^+e^-$ collisions at $\sqrt{s} = 2$ TeV.

\item[Fig.\ 4: \hfill]
Ratios of the convolutions of single-vector-boson distributions Eq.\
(\ref{conv}) and the exact luminosities for $pol=TT$, $\overline{TT}$,
$LT$ and $LL$ for a $W^+W^-$ pair in $e^+e^-$ collisions at $\sqrt{s} =
2$ TeV.

\item[Fig.\ 5: \hfill]
Ratios of the convolutions of single-vector-boson distributions Eq.\
(\ref{conv}) and the exact luminosities for $pol=TT$, $\overline{TT}$,
$LT$ and $LL$ for a $W^+W^-$ pair in $e^+e^-$ collisions at $\sqrt{s} =
4$ TeV.

\item[Fig.\ 6: \hfill]
Ratios of the leading logarithmic approximation for vector-boson pair
luminosities Eq.\ (\ref{lelo}) and the exact luminosities for
$pol=TT$, $\overline{TT}$, $LT$ and $LL$ for a $W^+W^-$ pair in
$e^+e^-$ collisions at $\sqrt{s} = 2$ TeV.

\item[Fig.\ 7: \hfill]
Luminosities ${\cal L}_{TT}({\cal W})$, ${\cal L}_{\overline{TT}}({\cal
W})$, ${\cal L}_{LT}({\cal W})$, and ${\cal L}_{LL}({\cal W})$ for a
$W^+W^-$ pair in $e^+e^-$ collisions at $\sqrt{s} = 500$ GeV.

\item[Fig.\ 8: \hfill]
Luminosities ${\cal L}_{TLTL}({\cal W})$, ${\cal
L}_{\overline{T}L\overline{T}L}({\cal W})$, and ${\cal
L}_{TTTT}({\cal W})$ for a $W^+W^-$ pair in $e^+e^-$ collisions at
$\sqrt{s} = 500$ GeV.

\item[Fig.\ 9: \hfill]
Luminosities ${\cal L}_{TT}({\cal W})$, ${\cal L}_{\overline{TT}}({\cal
W})$, ${\cal L}_{LT}({\cal W})$, and ${\cal L}_{LL}({\cal W})$ for a
$ZZ$ pair in $e^+e^-$ collisions at $\sqrt{s} = 500$ GeV.

\item[Fig.\ 10: \hfill]
Luminosities ${\cal L}_{TLTL}({\cal W})$,
${\cal L}_{\overline{T}L\overline{T}L}({\cal W})$, and ${\cal
L}_{TTTT}({\cal W})$ for a $ZZ$ pair in $e^+e^-$ collisions at
$\sqrt{s} = 500$ GeV.
\end{list}
\newpage

\begin{figure}
\unitlength1cm
\begin{picture}(15,22)
\put(0.,0.){\includegraphics{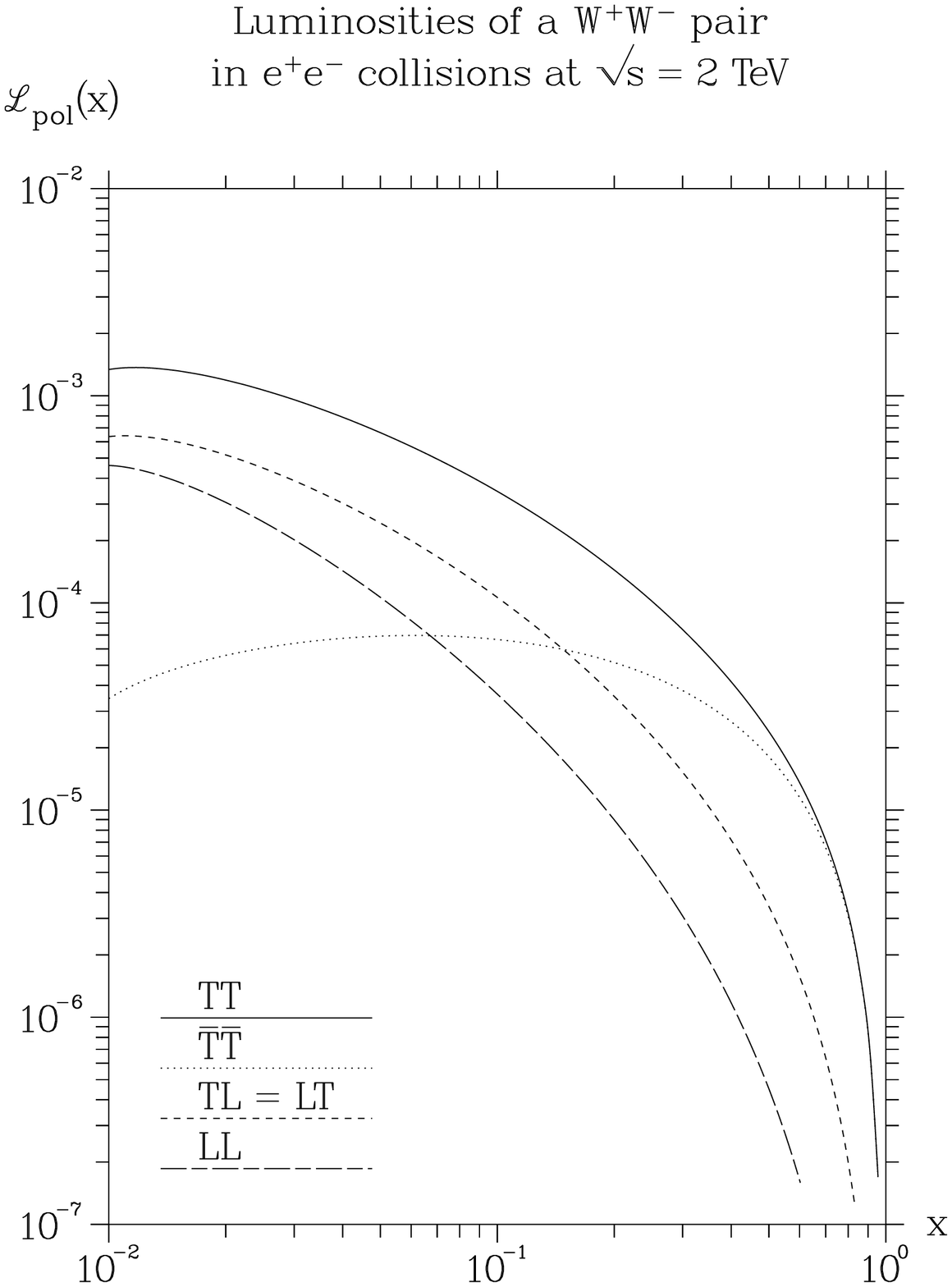}}
\end{picture}
\label{fig2}
\end{figure}
\noindent
Figure 2: Luminosities ${\cal L}_{TT}(x)$,
${\cal L}_{\overline{TT}}(x)$,
${\cal L}_{LT}(x)$, and ${\cal L}_{LL}(x)$ for a $W^+W^-$ pair in
$e^+e^-$ collisions at $\sqrt{s}=2$ TeV.

\newpage

\begin{figure}
\unitlength1cm
\begin{picture}(15,22)
\put(0,0){\includegraphics{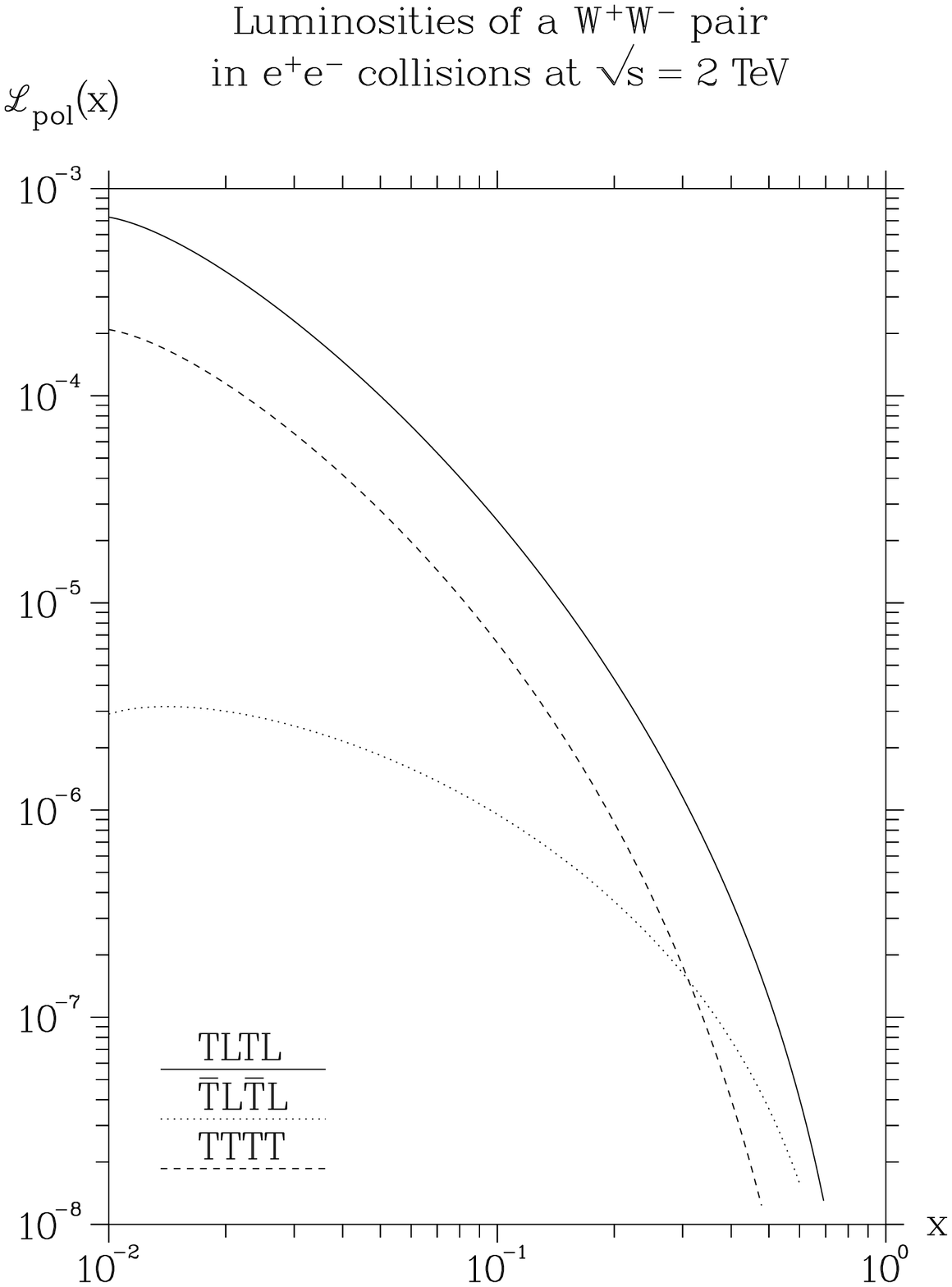}}
\end{picture}
\label{fig3}
\end{figure}
\noindent
Figure 3: Luminosities ${\cal L}_{TLTL}(x)$,
${\cal L}_{\overline{T}L\overline{T}L}(x)$, and ${\cal L}_{TTTT}(x)$
for a $W^+W^-$ pair in $e^+e^-$ collisions at $\sqrt{s} = 2$ TeV.

\newpage

\begin{figure}
\unitlength1cm
\begin{picture}(15,22)
\put(-0,0){\includegraphics{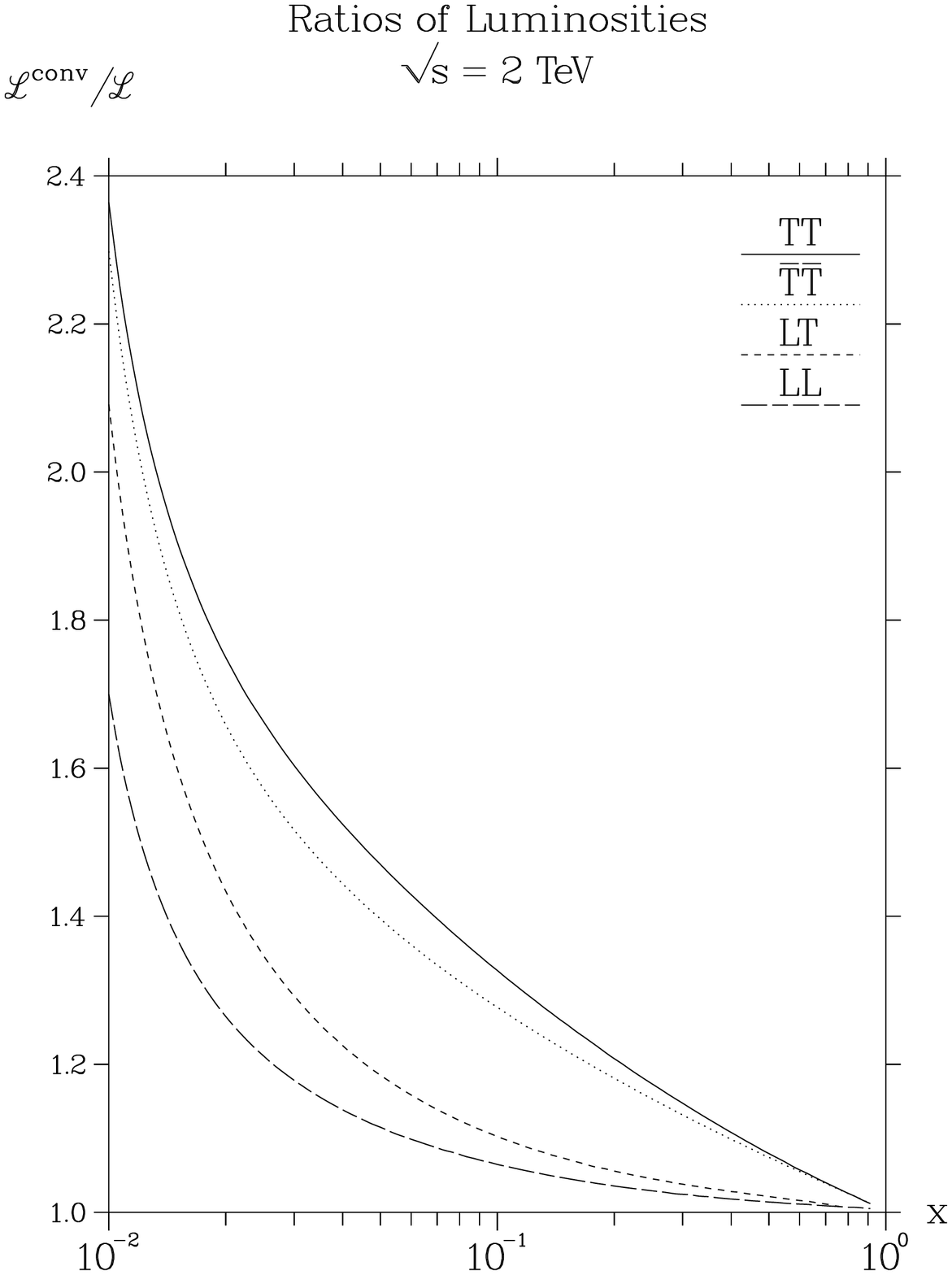}}
\end{picture}
\label{fig4}
\end{figure}
\noindent
Figure 4: Ratios of the convolutions of single-vector-boson
distributions Eq.\ (\ref{conv}) and the exact luminosities for
$pol=TT$, $\overline{TT}$, $LT$ and $LL$ for a $W^+W^-$ pair in
$e^+e^-$ collisions at $\sqrt{s} = 2$ TeV.

\newpage

\begin{figure}
\unitlength1cm
\begin{picture}(15,22)
\put(0,0){\includegraphics{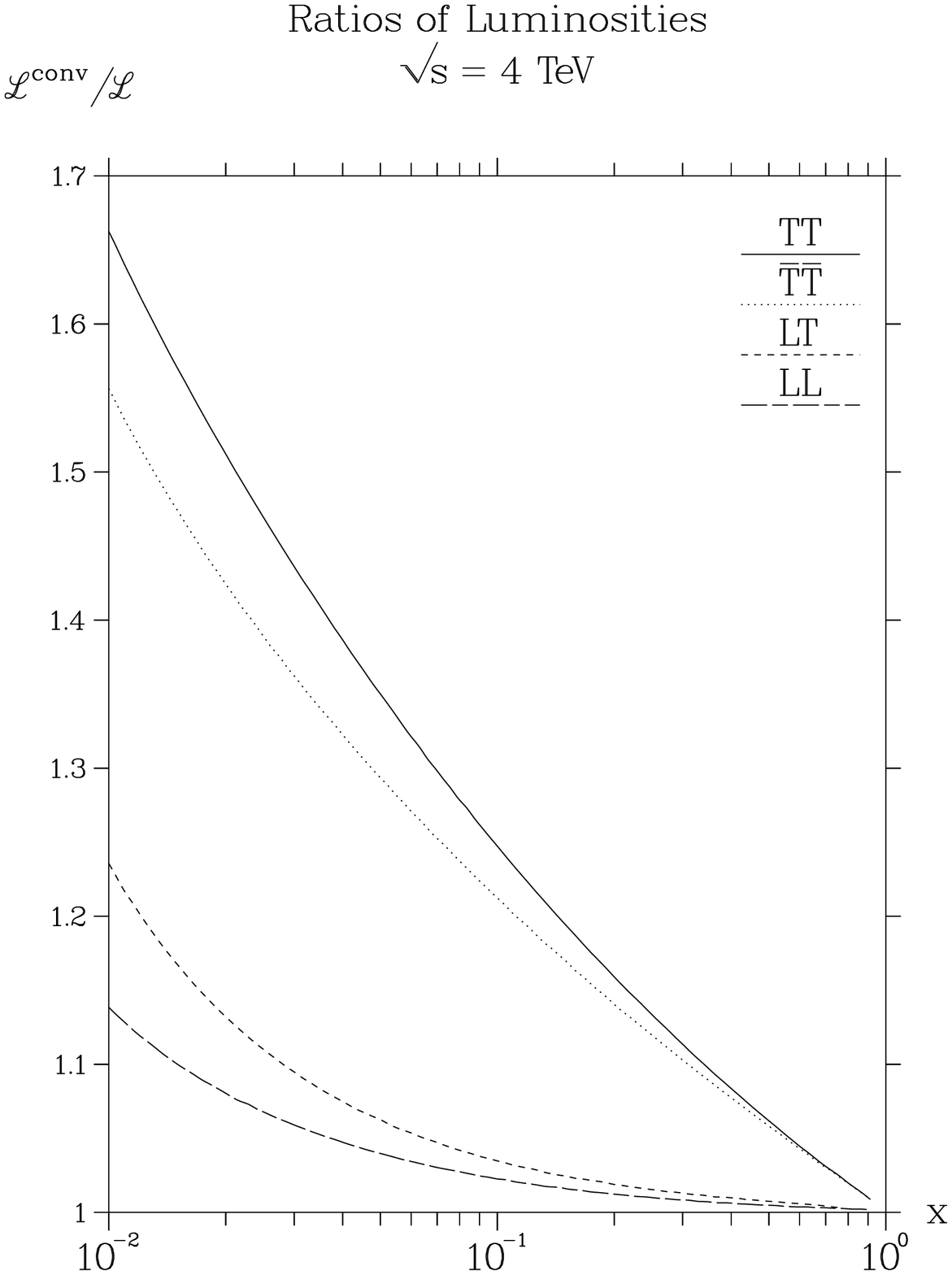}}
\end{picture}
\label{fig5}
\end{figure}
\noindent
Figure 5: Ratios of the convolutions of single-vector-boson
distributions Eq.\ (\ref{conv}) and the exact luminosities for
$pol=TT$, $\overline{TT}$, $LT$ and $LL$ for a $W^+W^-$ pair in
$e^+e^-$ collisions at $\sqrt{s} = 4$ TeV.

\newpage

\begin{figure}
\unitlength1cm
\begin{picture}(15,22)
\put(0,0){\includegraphics{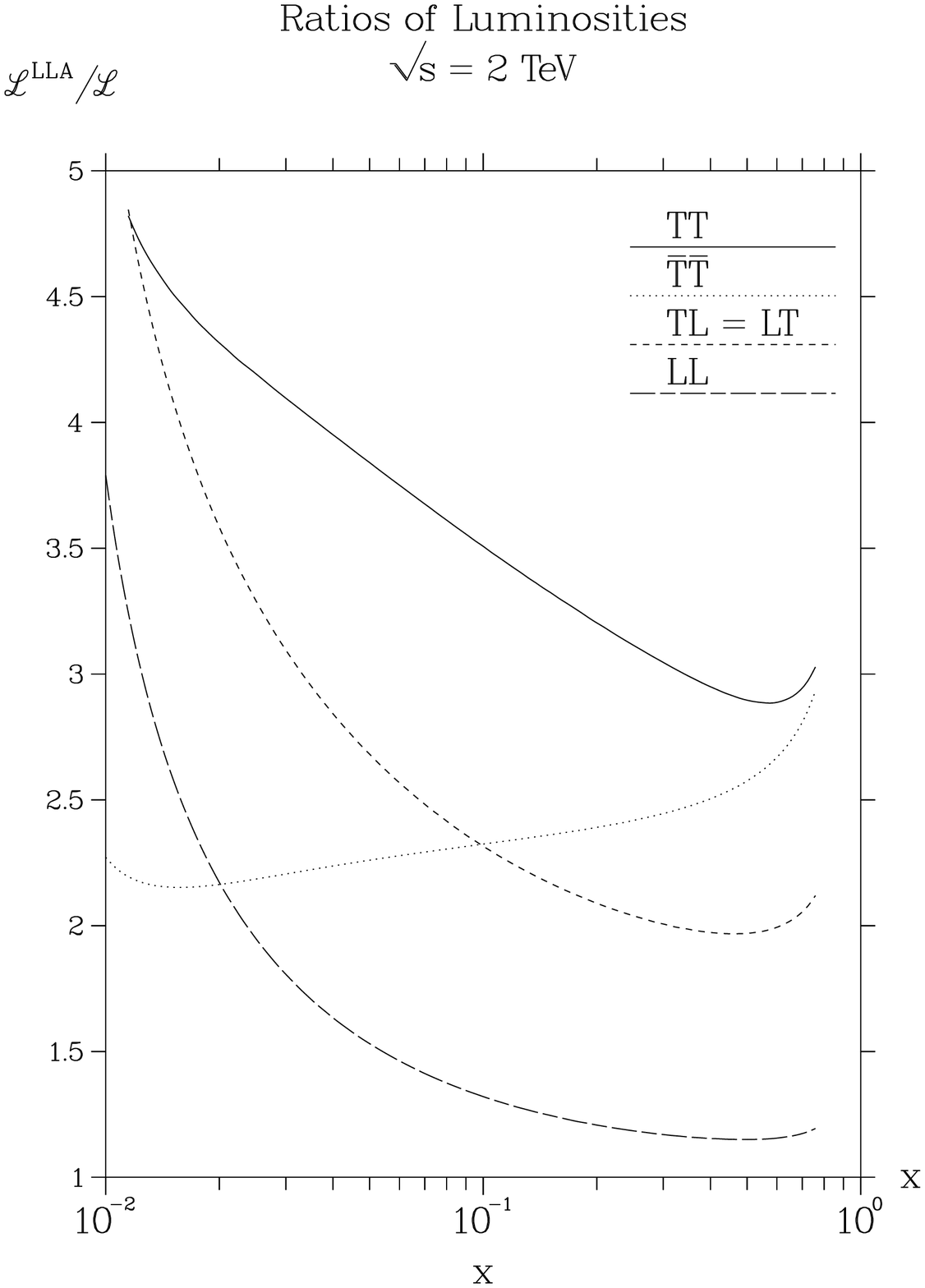}}
\end{picture}
\label{fig6}
\end{figure}
\noindent
Figure 6: Ratios of the leading logarithmic approximation for
vector-boson pair luminosities Eq.\ (\ref{lelo}) and the exact
luminosities for $pol=TT$, $\overline{TT}$, $LT$ and $LL$ for a
$W^+W^-$ pair in $e^+e^-$ collisions at $\sqrt{s} = 2$ TeV.

\newpage

\begin{figure}
\unitlength1cm
\begin{picture}(15,22)
\put(0,0){\includegraphics{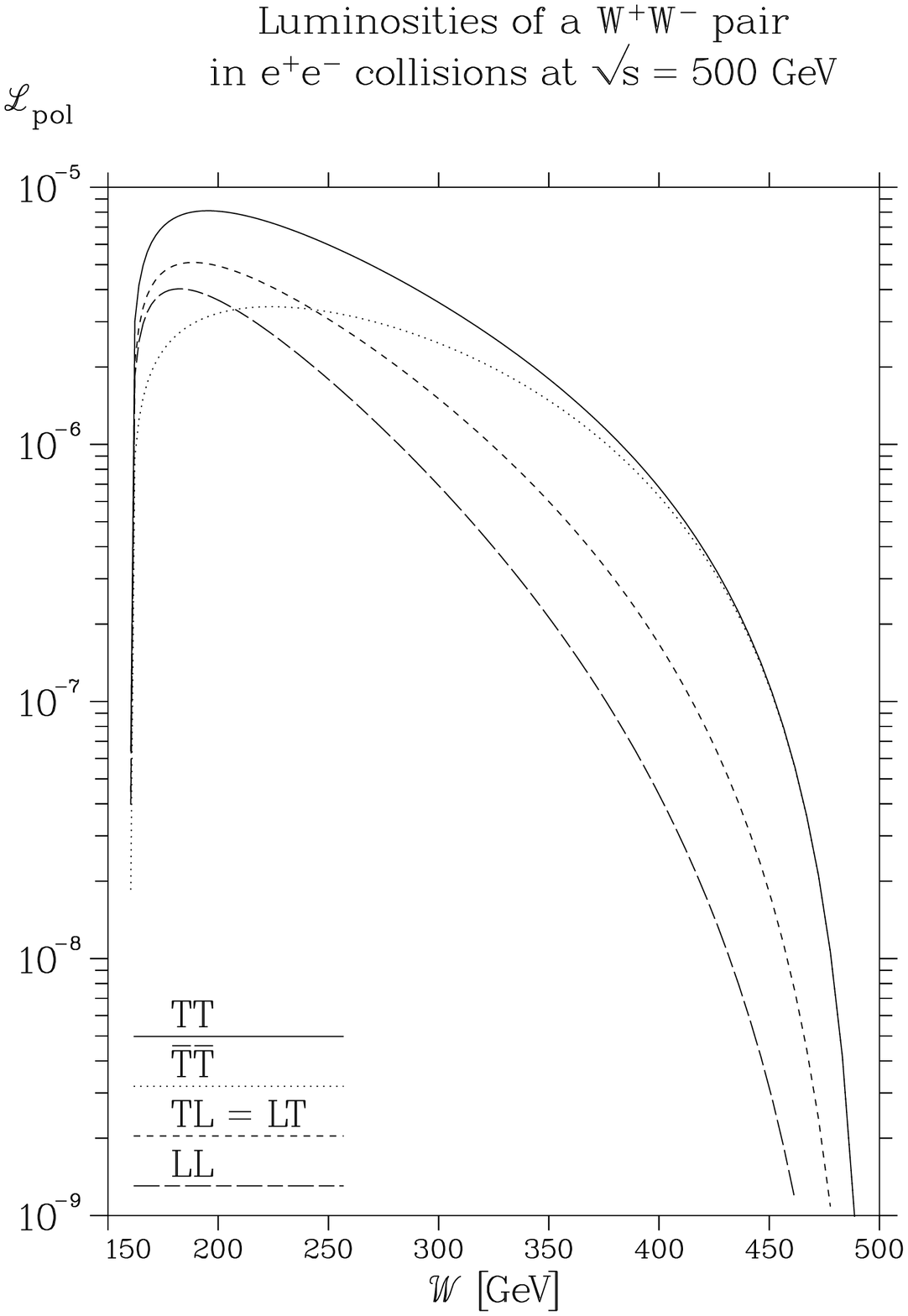}}
\end{picture}
\label{fig7}
\end{figure}
\noindent
Figure 7: Luminosities ${\cal L}_{TT}$, ${\cal L}_{\overline{TT}}$,
${\cal L}_{LT}$, and ${\cal L}_{LL}$ as a function of the boson pair
invariant mass ${\cal W}$, ${\cal W}^2 =xs$, for a $W^+W^-$ pair in
$e^+e^-$ collisions at $\sqrt{s} = 500$ GeV.

\newpage

\begin{figure}
\unitlength1cm
\begin{picture}(15,22)
\put(0,0){\includegraphics{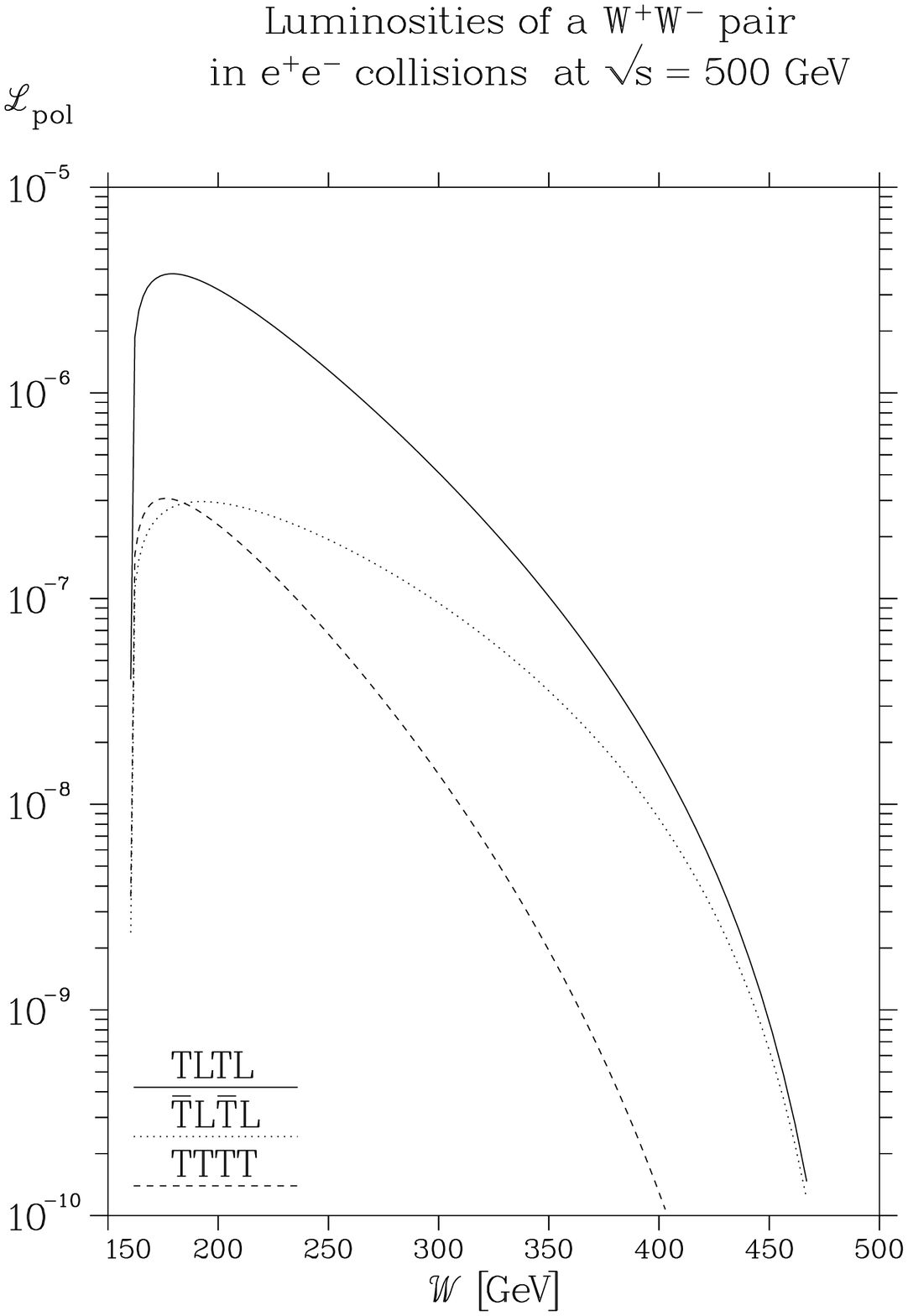}}
\end{picture}
\label{fig8}
\end{figure}
\noindent
Figure 8: Luminosities ${\cal L}_{TLTL}$, ${\cal
L}_{\overline{T}L\overline{T}L}$, and ${\cal L}_{TTTT}$ as a function
of the boson pair invariant mass ${\cal W}$, ${\cal W}^2 =xs$, for a
$W^+W^-$ pair in $e^+e^-$ collisions at $\sqrt{s} = 500$ GeV.

\newpage

\begin{figure}
\unitlength1cm
\begin{picture}(15,22)
\put(0,0){\includegraphics{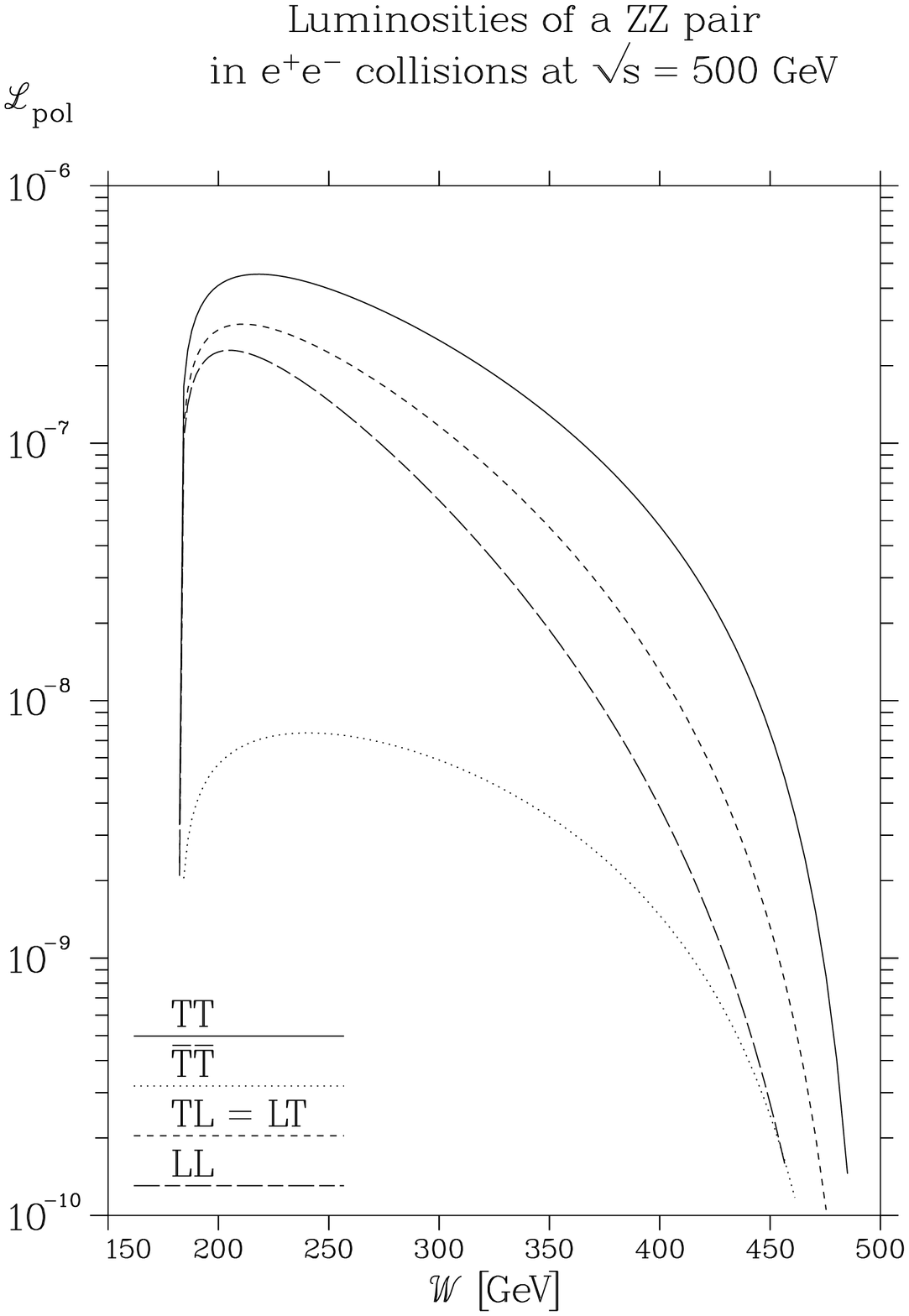}}
\end{picture}
\label{fig9}
\end{figure}
\noindent
Figure 9: Luminosities ${\cal L}_{TT}$, ${\cal L}_{\overline{TT}}$,
${\cal L}_{LT}$, and ${\cal L}_{LL}$ as a function of the boson pair
invariant mass ${\cal W}$, ${\cal W}^2 =xs$, for a $ZZ$ pair in
$e^+e^-$ collisions at $\sqrt{s} = 500$ GeV.

\newpage

\begin{figure}
\unitlength1cm
\begin{picture}(15,22)
\put(0,0){\includegraphics{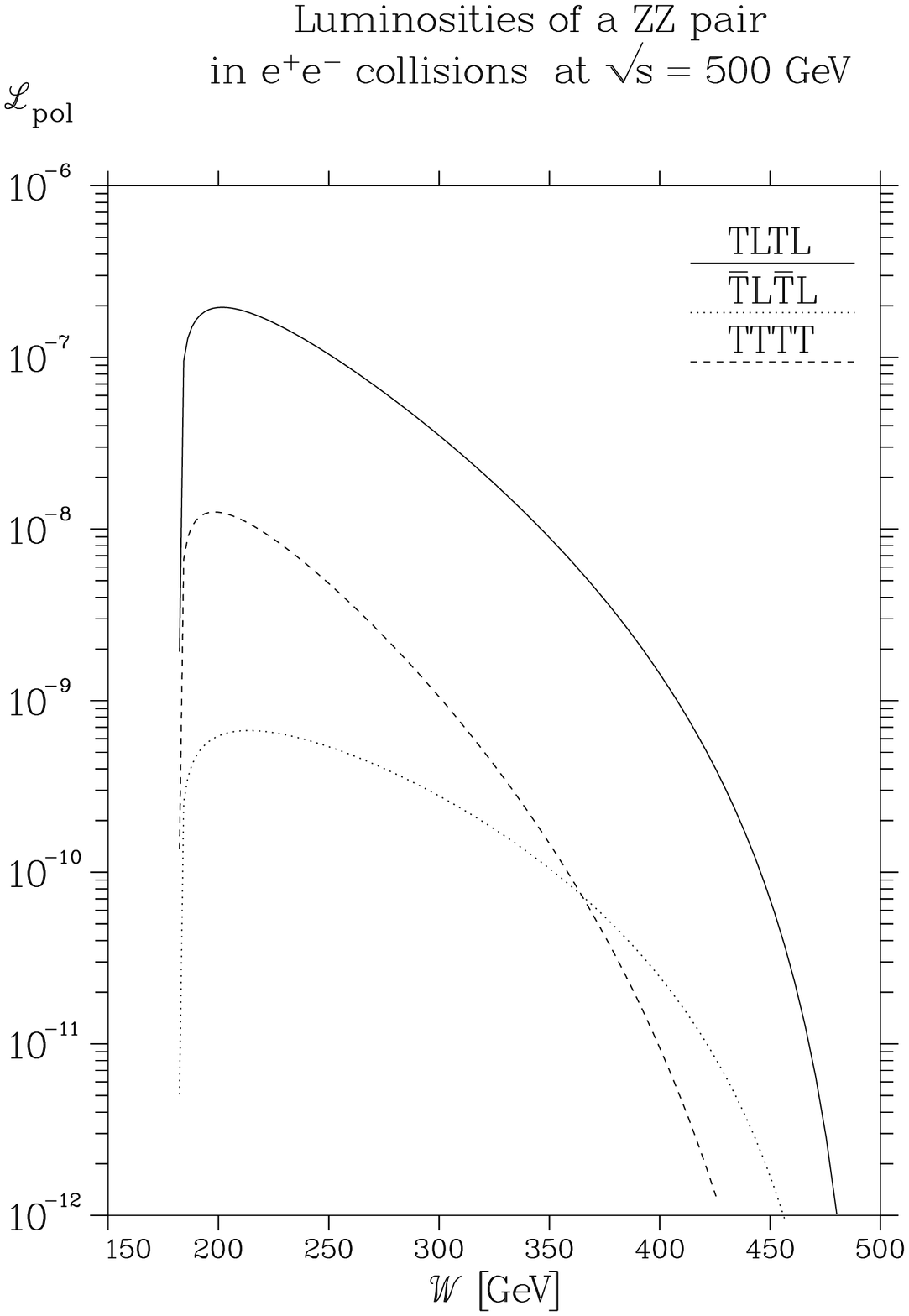}}
\end{picture}
\label{fig10}
\end{figure}
\noindent
Figure 10: Luminosities ${\cal L}_{TLTL}$, ${\cal
L}_{\overline{T}L\overline{T}L}$, and ${\cal L}_{TTTT}$ as a function
of the boson pair invariant mass ${\cal W}$, ${\cal W}^2 =xs$, for a
$ZZ$ pair in $e^+e^-$ collisions at $\sqrt{s} = 500$ GeV.


\begin{thebibliography}{99}

\bibitem{WWA}
E.\ Fermi, Z.\ Phys.\ 29 (1924) 315; \\
C.\ Weizs\"acker, Z.\ Phys.\ 88 (1934) 612; \\
E.\ Williams, Phys.\ Rev.\ 45 (1934) 729.

\bibitem{kane1}
G.\ L.\ Kane, Proc.\ "Physics of the XXIst century", Tucson, Arizona,
Dec.\ 1983.

\bibitem{dawson}
S.\ Dawson, Nucl.\ Phys.\ B249 (1985) 42.

\bibitem{kane2}
G.\ L.\ Kane, W.\ W.\ Repko, and W.\ B.\ Rolnick, Phys.\ Lett.\ 148B
(1984) 367.

\bibitem{lindfors}
J.\ Lindfors, Z.\ Phys.\ C28 (1985) 427.

\bibitem{cahn}
R.\ Cahn and S.\ Dawson, Phys.\ Lett.\ 136B (1984) 196; Err.\ ibid.\
138B (1984) 464.

\bibitem{HIGGS}
R.\ Cahn, Nucl.\ Phys.\ B255 (1985) 341; Err.\ ibid.\ B262 (1985) 744;
\\
D.\ A.\ Dicus, S.\ Willenbrock, Phys.\ Rev.\ D32 (1985) 1642;
\\
M.\ J.\ Duncan, G.\ L.\ Kane and W.\ W.\ Repko, Nucl.\ Phys.\ B272
(1986) 517;
\\
G.\ Altarelli, B.\ Mele, F.\ Pitalli, Nucl.\ Phys.\ B287 (1987) 205;
\\
M.\ C.\ Bento and C.-H.\ Llewellyn-Smith, Nucl.\ Phys.\ B289 (1987) 36;
\\
D.\ A.\ Dicus, K.\ J.\ Kallianpur, and S.\ D.\ Willenbrock, Phys.\
Lett.\ B200 (1988) 187.

\bibitem{HEAVYQ}
S.\ Willenbrock and D.\ A.\ Dicus, Phys.\ Rev.\ D34 (1986) 155;
\\
J.\ Lindfors, Z.\ Phys.\ C33 (1987) 385;
\\
S.\ Dawson, G.\ L.\ Kane, C.\ P.\ Yuan, and S.\ Willenbrock, Proc.\ of
the 1986 Summer Study on Physics of the Superconducting Super Collider,
Snowmass, CO, Jun 23 -- Jul 11, 1986, p.\ 235;
\\
S.\ Dawson and S.\ Willenbrock, Nucl.\ Phys.\ B284 (1987) 449;
\\
R.\ P.\ Kauffman, Phys.\ Rev.\ D41 (1990) 3343.

\bibitem{VVscatt}
M.\ Chanowitz and M.\ K.\ Gaillard, Phys.\ Lett.\ 142B (1984) 85; \\
M.\ Chanowitz and M.\ K.\ Gaillard, Nucl.\ Phys.\ B261 (1985) 379;
\\
J.\ F.\ Gunion, J.\ Kalinowski, A.\ Tofighi-Niaki, A.\ Abbasabadi, and
W.\ Repko, Proc.\ of the 1986 Summer Study on Physics of the
Superconducting Super Collider, Snowmass, CO, Jun 23 -- Jul 11, 1986,
p.\ 156;
\\
J.\ F.\ Gunion, J.\ Kalinowski, and A.\ Tofighi-Niaki, Phys.\ Rev.\
Lett.\ 57 (1986) 2351;
\\
B.\ Mele, in Proc.\ of La Thuile Workshop on Physics at Future
Accelerators, La Thuile, Italy, Jan 7 -- 13, 1987;
\\
A.\ Abbasabadi, W.\ W.\ Repko, D.\ A.\ Dicus, and R.\ Vega, Phys.\ Rev.\
D38 (1988) 2770;
\\
D.\ A.\ Dicus, S.\ L.\ Wilson, and R.\ Vega, Phys.\ Lett.\ B192 (1987)
231;
\\
M.\ Kuroda, F.\ M.\ Renard and D.\ Schildknecht, Z.\ Phys.\ C40 (1988)
575;
\\
G.\ J.\ Gounaris and F.\ M.\ Renard, Z.\ Phys.\ C59 (1993) 143.

\bibitem{dicus1}
D.\ A.\ Dicus and R.\ Vega, Phys.\ Rev.\ Lett.\ 57 (1986) 1110.

\bibitem{TRANSV}
A.\ Abbasabadi and W.\ W.\ Repko, Proc.\ of the 1986 Summer Study on
Physics of the Superconducting Super Collider, Snowmass, CO, Jun 23 --
Jul 11, 1986, p.\ 154;
\\
J.\ P.\ Ralston and F.\ Olness, Proc.\ of the 1986 Summer Study on
Physics of the Superconducting Super Collider, Snowmass, CO, Jun 23 --
Jul 11, 1986, p.\ 191;
\\
A.\ Abbasabadi and W.\ W.\ Repko, Phys.\ Rev.\ D36 (1987) 289;
\\
A.\ Abbasabadi and W.\ W.\ Repko, Phys.\ Rev.\ D50 (1994) 5704;
\\
W.\ W.\ Repko and W.-K.\ Tung, Proc.\ of the 1986 Summer Study on
Physics of the Superconducting Super Collider, Snowmass, CO, Jun 23 --
Jul 11, 1986, p.\ 159.

\bibitem{tung1}
P.\ W.\ Johnson, F.\ I.\ Olness, and Wu-Ki Tung, Proc.\ of the 1986
Summer Study on Physics of the Superconducting Super Collider,
Snowmass, CO, Jun 23 -- Jul 11, 1986, p.\ 164; \\
P.\ W.\ Johnson, F.\ I.\ Olness, and Wu-Ki Tung, Phys.\ Rev.\ D36 (1987)
291.

\bibitem{lindfors3}
J.\ Lindfors, Z.\ Phys.\ C35 (1987) 355.

\bibitem{renard}
M.\ Capdequi Peyran\'{e}re, J.\ Layssac, H.\ Leveque, G.\ Moultaka and
F.\ M.\ Renard, Z.\ Phys.\ C41 (1988) 99.

\bibitem{dobrov}
A.\ Dobrovolskaia and V.\ Novikov, preprint LPTHE 93/14 (April 1993).

\bibitem{budnev}
V.\ M.\ Budnev, I.\ F.\ Ginzburg, G.\ V.\ Meledin, and V.\ G.\ Serbo,
Phys.\ Rep.\ 15C (1974) 183; \\
N.\ S.\ Craigie, K.\ Hidaka, M.\ Jacob and F.\ M.\ Renard, Phys.\ Rep.\
99 (1983) 69.

\bibitem{bonneau}
G.\ Bonneau, M.\ Gourdin, and F.\ Martin, Nucl.\ Phys.\ B54 (1973) 573;
\\
G.\ Bonneau and F.\ Martin, Nucl.\ Phys.\ B68 (1974) 367.

\bibitem{kleiss}
R.\ Kleiss and J.\ W.\ Stirling, Phys.\ Lett.\ B182 (1986) 75.

\bibitem{kunszt}
Z.\ Kunszt and D.\ E.\ Soper, Nucl.\ Phys.\ B296 (1988) 253.

\bibitem{cortese}
S.\ Cortese and R.\ Petronzio, Phys.\ Lett.\ B276 (1992) 203.

\bibitem{godbole}
R.\ M.\ Godbole and F.\ Olness, Int.\ J.\ Mod.\ Phys.\ A2 (1987) 1025;
\\
R.\ M.\ Godbole and S.\ D.\ Rindani, Phys.\ Lett.\ B190 (1987) 192; Z.\
Phys. C36 (1987) 395.

\end{thebibliography}
\end{document}